\documentclass[reprint,twocolumn,amsmath,amssymb,aps,nofootinbib,longbibliography]{revtex4-1}
\usepackage{tablefootnote}
\usepackage[T1]{fontenc}
\usepackage{amsmath}
\usepackage{amssymb}
\usepackage{adjustbox}
\usepackage{multirow}
\usepackage{hyperref}
\usepackage{bbm}
\usepackage{xcolor}
\usepackage{braket}
\usepackage{slashed}
\usepackage{diagbox}
\usepackage[utf8]{inputenc}
\usepackage{tcolorbox}
\usepackage{tabularx}
\usepackage{bbm,bm}
\usepackage[normalem]{ulem}
\usepackage[splitrule,bottom]{footmisc} % <-- new

\newcommand{\yong}[1]{{\color{black}#1}}
\newcommand{\abs}[1]{\left\vert #1 \right\vert}
\newcommand{\eqal}[1]{\begin{align}#1\end{align}}
\newcommand{\bewt}[1]{\begin{widetext}#1\end{widetext}}
\newcommand{\eqsp}[1]{\begin{equation}\begin{split} #1\end{split}\end{equation}}
\newcommand{\nb}{\nonumber}
\newcommand{\bdsb}[1]{\boldsymbol{#1}}
\newcommand{\p}[1]{$\mathbf{P}$}
\newcommand{\cp}[1]{$\mathbf{CP}$}

\newcommand{\hatp}{\hat{\bm p}}
\newcommand{\hatk}{\hat{\bm k}}

\newcommand{\hatl}{\hat{\bm l}}

\begin{document}
\raggedbottom

\title{Fundamental Tests of P and CP Symmetries\\ Using Octet Baryons at the $J/\psi$ Threshold}

\author{Yong Du}
\email{yongdu5@sjtu.edu.cn}
\affiliation{Tsung-Dao Lee Institute, Shanghai Jiao Tong University, Shanghai 200240, China}

\author{Xiao-Gang He}
\email{hexg@sjtu.edu.cn}
\affiliation{Tsung-Dao Lee Institute, Shanghai Jiao Tong University, Shanghai 200240, China}
\affiliation{Key Laboratory for Particle Astrophysics and Cosmology (MOE) \& Shanghai Key Laboratory for Particle Physics and Cosmology, Shanghai Jiao Tong University, Shanghai 200240, China}

\author{Jian-Ping Ma}
\email{majp@itp.ac.cn}
\affiliation{CAS Key Laboratory of Theoretical Physics, Institute of Theoretical Physics, P.O. Box 2735, Chinese Academy of Sciences, Beijing 100190, China}
\affiliation{School of Physical Sciences, University of Chinese Academy of Sciences, Beijing 100049, China}
\affiliation{School of Physics and Center for High-Energy Physics, Peking University, Beijing 100871, China}

\author{Xin-Yu Du}
\email{2020dxy@sjtu.edu.cn}
\affiliation{Tsung-Dao Lee Institute, Shanghai Jiao Tong University, Shanghai 200240, China}
\affiliation{Key Laboratory for Particle Astrophysics and Cosmology (MOE) \& Shanghai Key Laboratory for Particle Physics and Cosmology, Shanghai Jiao Tong University, Shanghai 200240, China}

\date{\today}

\begin{abstract}
We {systematically} investigate tests of the parity P and the combined parity and charge-conjugate CP symmetries from differential angular distributions of $J/\psi$ decaying into the lowest-lying baryon pairs at BESIII and the next-generation super tau-charm facilities (STCFs).
Large corrections from $Z$ and $W$ exchange induced parity violating effects are found for $J/\psi$ decays with large logarithms resummed up to $\mathcal{O}(\alpha_s)$.
The parity-violating asymmetries on the production and the decay sides of $J/\psi$ are both {estimated} to be of $\mathcal{O}(10^{-4})$, thus barely observable with the 10 billion $J/\psi$ events currently collected at BESIII. Nevertheless, these asymmetries utilizing the current BESIII data already permit a {measurement} of the weak mixing angle with an absolute uncertainty $\delta s_w^2\approx{0.09}$, corresponding to the first determination of $s_w^2$ at the $J/\psi$ threshold. {In the future,} STCFs are estimated to improve this {bound} by a factor of $\sim\,20$ {to $\delta s_w^2\approx{0.005}$} within one year based on luminosity rescaling.
We also obtain the 95\% confidence level upper bounds on the electric dipole moments of the octet baryons, which are of $\mathcal{O}(10^{-18})\,e{\rm\,cm}$ for BESIII and $\mathcal{O}(10^{-19})\,e{\rm\,cm}$ for STCFs. These bounds are improved by two to three orders of magnitude in comparison with the only existing one on $\Lambda$ from Fermilab. The method discussed in this work also paves a way for a first and direct measurement of the $\Xi$ and $\Sigma$ electric dipole moments.
\end{abstract}

\maketitle

%%%%%%%%%%%%%%%%%%%%%
\section{Introduction}\label{sec:intro}
%%%%%%%%%%%%%%%%%%%%%
Symmetries, along with the tests in terms of their violation and/or conservation at different energies and processes, have played an essential role in understanding the laws of nature. Starting with the pioneering work by Lee and Yang\,\cite{Lee:1956qn} in 1956, the conservation of parity P in beta, hyperon and meson decays was questioned, whose violation was confirmed one year later by the now widely known Wu experiment using polarized Co$^{60}$\,\cite{Wu:1957my} and also soon confirmed in pion and muon decays\,\cite{Garwin:1957hc,Friedman:1957mz}. The breakthrough in this regard, on the one hand, opened a completely new chapter in physics, and on the other hand, has been driving physicists in questioning the conservation of the other fundamental symmetries in physics such as the time reversal and the charge-conjugate symmetries\,\cite{Lee:1957qq}.

The dedicated efforts in this regard had led to the first observation of the violation of the combined symmetry of charge-conjugate and parity CP at Brookhaven\,\cite{Christenson:1964fg} in 1964, through the 2$\pi$ decay mode of neutral Kaons. Subsequent studies at SLAC from the Babar\,\cite{BaBar:2001pki} and at KEK from the BELLE\,\cite{Belle:2001zzw} collaborations also revealed the same phenomenology with the $B^0$ meson. Recently the LHCb collaboration also observed this violation in charm decays with the $D^0$ meson\,\cite{LHCb:2019hro}.

Theoretically, the lasting and keen interest in CP violation can be traced back to the observed baryon-antibaryon asymmetry of the Universe\,\cite{Hu:2001bc,Cyburt:2004cq,Planck:2018nkj}, rooted in the violation of this combined symmetry as one of the necessary conditions from Sakharov\,\cite{Sakharov:1967dj}. It has been well-known that the established and well tested Standard Model (SM) of particle physics is insufficient in producing the aforementioned asymmetry of our Universe\,\cite{Cline:2006ts}, thus new sources of CP violation beyond the SM have to be identified along with the widely studied meson systems presented in the last paragraph.

The weak decay of hyperons provides such an alternative. Theoretically, using the branching ratios of different decay modes of $\Sigma^\pm$, the observation of CP violation was firstly investigated by Okubo in 1958\,\cite{Okubo:1958zza}. Soon after, a short discussion on parity and CP violation for anti-baryon interactions was presented in\,\cite{Pais:1959zza}. A systematical study on CP violation using the rate differences between particles and antiparticles of the octet baryons was performed in\,\cite{Brown:1983wd}, where upper bounds within the quark model were estimated. Using a different approach, ref.\,\cite{Chau:1983ei} explored the possibility of CP violation measurements from the angular distributions of non-leptonic decay modes of the octet baryons, with a systematic study on all the nonleptonic decay modes of the hyperons performed in{\,\cite{Donoghue:1985ww,Donoghue:1986hh}}. In the context of new physics where non-standard semileptonic interactions exist, the SM Effective Field Theory (SMEFT) or its low-energy variants for example, hyperon decays have been studied in\,\cite{Chang:2014iba} in combination with data from the Large Hadron Collider (LHC) at CERN. See also\,\cite{Li:2016tlt,Goudzovski:2022vbt} and references therein in this context.

Experimentally, searches of CP violation with hyperons had been carried out at Fermilab in 1981\,\cite{Pondrom:1981gu}, where the only existing electric dipole moment result for the $\Lambda$ baryon was reported. Studies on CP violation were later also carried out by the R608 collaboration at CERN, where the $\Lambda$ baryon as a target of CP violation was produced from $p\bar p$ collision at $\sqrt s=30.8\rm\,GeV$\,\cite{R608:1985fmh} and no CP violation was observed. Subsequent studies at DCI Orsay from the DM2 collaboration further investigated the hyperon system by studying $J/\psi\to\Lambda\bar\Lambda$\,\cite{DM2:1988ppi} and $J/\psi\to p\bar p,\,\Lambda\bar\Lambda,\,\Sigma^0\bar\Sigma^0$, where the CP parameters\,\cite{Lee:1957qs} of these baryons were reported in\,\cite{DM2:1987riy}. CP violation in the $\Xi^-$ baryon was carried out by the HyperCP collaboration at Fermilab, where the hyperons were produced with an 800\,GeV $p$ beam on the copper target. The produced hyperons were then used to measure the CP-violating asymmetries in $\Xi^-\to\Lambda\pi^-\to p\pi^-\pi^-$ and $\bar\Xi^-\to\bar\Lambda\pi^+\to\bar p\pi^+\pi^+$ by the E871 experiment\,\cite{HyperCP:2004zvh}. However, no CP violation was observed due to limited statistics, as well as very large systematical uncertainties. At higher energies, the LHCb\,\cite{LHCb:2016yco} recently reported the evidence of CP violation at the $3.3\sigma$ level from beauty baryon decays of $\Lambda_b^0\to p\pi^-\pi^+\pi^-$. On the other hand, with the very rich events accumulated at BESIIII at the $J/\psi$ mass threshold, the recent analysis on $J/\psi\to\Lambda\bar\Lambda$ from the BESIII collaboration shows no signal of CP violation and also presents very precise decay parameters for the $\Lambda$ that are dramatically different from the world average at that time\,\cite{BESIII:2018cnd}. These results from BESIII further modify the other experiments using the proton to reconstruct $\Lambda$, refs.\,\cite{Bunce:1976yb,Pondrom:1985aw,FNALE756:2003kkj,ATLAS:2014ona,STAR:2017ckg} for example, and are also favored by an independent analysis from the CLAS collaboration\,\cite{Ireland:2019uja}.

Using the $\Lambda$ baryon and the other lowest-lying baryons for testing the CP symmetry has been further investigated at BESIII since then\,\cite{BESIII:2019nep,BESIII:2019cuv,BESIII:2021ypr,BESIII:2021cvv,BESIII:2022qax,BESIII:2022lsz,BESIII:2023sgt,BESIII:2023cvk}, and theoretically in\,\cite{He:1992ng,He:2022jjc} for $J/\psi\to\Lambda\bar\Lambda$ and in\,\cite{Hong:2023soc} for $J/\psi\to\Xi^0\bar\Xi^0$. In view of this progress and the tens of billions $J/\psi$ particles already collected at BESIII, we think it timely to systematically investigate tests of the parity and the CP symmetries with $J/\psi$ decay. This study is further motivated by the STCFs\,\cite{Charm-TauFactory:2013cnj,Achasov:2023gey} that is expected to deliver $3.4\times10^{12}$ $J/\psi$ particles at its mass threshold \textit{each year}. The rich physics of $J/\psi$ has been reviewed in\,\cite{Kopke:1988cs}, and for this work, we only focus on $J/\psi$ produced at rest at BESIII and STCFs that subsequently decays into the lowest-lying baryons for testing the parity and the CP symmetries. Furthermore, we only consider the unpolarized beam option of these facilities and we refer the readers to \,\cite{Bondar:2019zgm,Zeng:2023wqw,Fu:2023ose,Cao:2024tvz} for recent studies with polarized beams. From this study, we find that:
\begin{itemize}
\item In calculating the parity violating hadronic form factors of $J/\psi$ decay, we find some generic modifications by a factor of a few when renormalization group evolved from the weak scale down to the decay scale in the SM. In particular, the axial-vector form factor of $\Sigma^0$ is found to be reduced by a factor of $\sim\,40$ as summarized in table\,\ref{tab:fanumres} for all the octet baryons;
\item The parity-violating asymmetries introduced in this work are {evaluated} to be of $\mathcal{O}(10^{-4})$, as seen in table\,\ref{tab:pcpvio}, which are slightly below the statistical uncertainties with the 10 billion $J/\psi$ particles collected up to now at BESIII. A combined analysis of these parity-violating asymmetries from both the production and the decay of $J/\psi$, on the other hand, already leads to a determination of the weak mixing angle with an absolute uncertainty of $\delta s_w^2\approx{0.09}$, corresponding to the first measurement at the mass pole of $J/\psi$, as summarized in table\,\ref{tab:swunc3} and shown in figure\,\ref{fig:swdet}. Based on luminosity rescaling, we also estimate that this {sensitivity} in $s_w^2$ is expected to be improved by about a factor of $\sim$20 at  STCFs within one year; 
\item The 95\% confidence level sensitivity for the upper bounds on the electric dipole moments of the octet baryons are found to be of $\mathcal{O}(10^{-18})\,e{\rm\,cm}$ at BESIII, about two orders of magnitude stronger than the only measured one on $\Lambda$ from Fermilab more than 40 years ago. These upper bounds from BESIII are expected to be further improved to $\mathcal{O}(10^{-19})\,e{\rm\,cm}$ at STCFs with just one-year data collection, as summarized in table\,\ref{tab:pcpvio}.
\end{itemize}

In the next sections, we provide details of our calculations. The rest of the paper is organized as follows: We briefly review the theoretical formalism for testing P and the CP symmetries with threshold $J/\psi$ production and decay in section\,\ref{sec:setup}, based on which we define the asymmetric observables that can be directly connected with experiments. These observables depend on hadronic form factors, whose derivations are detailed in section\,\ref{sec:formfac}, where their renormalization group evolution is discussed. The results from this study are presented in section\,\ref{sec:res}, and we draw our conclusions in section\,\ref{sec:con}.

%%%%%%%%%%%%%%%%%%%%%
\section{Theoretical setup}\label{sec:setup}
%%%%%%%%%%%%%%%%%%%%%
As mentioned in the Introduction, the abundant production of $J/\psi$ at BESIII and STCFs makes $J/\psi$ an ideal testbed for fundamental symmetries such as P and the CP symmetries. In this section, we discuss how to test these fundamental symmetries in  $J/\psi$ decaying into the baryon-antibaryon pairs $B \bar B$ with $B=\Lambda,\Sigma^{\pm,0},\Xi^{\pm,0}$, where $B$ or $\bar B$ subsequently undergoes two-body weak decays.  

We consider the  production of a $B\bar B$ system through the following process: 
\eqal{
e^-(p_1)+e^+(p_2) \to J/\psi \to B(k_1,s_1)+\bar B(k_2,s_2).
}
In the center-of-mass frame, we define the momenta of the initial leptons and the baryons as
\eqal{
p_1^\mu &\, = (E_c,{\bm p}),\quad p_2^\mu = (E_c,-{\bm p}),\nb\\
k_1^\mu &\, = (E_c,{\bm k}),\quad k_2^\mu = (E_c,-{\bm k}),
}
with $E_c = m_{J/\psi}/2$. $s_{1,2}$ are spin vectors of the final state particles in their respective rest frames as
\eqal{
s_1^\mu = (0, {\bm s_1}), \quad s_2^\mu = (0,{\bm s_2}).
}
In this work, we only consider the case where the initial electron and positron are unpolarized. However, the analysis can be directly generalized to the polarized case such as STCF\,\cite{Achasov:2023gey}, Belle-II\,\cite{Belle-II:2018jsg}, ILC\,\cite{ILC:2013jhg,ILCInternationalDevelopmentTeam:2022izu}, CLIC\,\cite{CLICPhysicsWorkingGroup:2004qvu,Robson:2018zje}, and CEPC\,\cite{CEPCStudyGroup:2023quu}. In this unpolarized scenario, the probability of this considered process can only depend on ${\bm p}$, ${\bm k}$ and ${\bm s_{1,2}}$:
\begin{equation} 
   R({\bm p}, {\bm k}, {\bm s_1}, {\bm s_2}) = \frac{1}{4}\sum  \vert \bra{f}\mathcal{T}\ket{i} \vert^2, 
\label{RJ}    
\end{equation}    
where $i$ and $f$ are particle collections in the initial and final states, respectively, and the sum is over the spins 
of the initial states. If P or CP is conserved, $R$ has the following properties:
\eqsp{
R ({\bm p}, {\bm k}, {\bm s_1}, {\bm s_2})  &\, = R (-{\bm p}, -{\bm k}, {\bm s_1}, {\bm s_2}),\\
R ({\bm p}, {\bm k}, {\bm s_1}, {\bm s_2})  &\, = R ({\bm p}, {\bm k}, {\bm s_2}, {\bm s_1}),\label{eq:cp2sfi}
}
where the first and the second equalities are from P and CP conservation, respectively. From these relations, it is required to measure the spins of final state particles for testing P and CP symmetries\,\cite{Bernreuther:1988jr}. The probability $R$ can be decomposed as
\eqsp{
 R({\bm p}, {\bm k}, {\bm s_1}, {\bm s_2}) = &\, a(\hatp\cdot\hatk) + {\bm s}_1 \cdot \bdsb{B_1}(\bdsb{\hat p},\bdsb{\hat k})\\
&\,  + {\bm s}_2 \cdot \bdsb{B_2}(\bdsb{\hat p},\bdsb{\hat k})- s_1^i s_2^j  C^{ij}(\bdsb{\hat p},\bdsb{\hat k}),\label{eq:spindocom}
} 
 where
 \begin{equation} 
     \hatp =\frac{{\bm p}}{\vert{\bm p}\vert}, \quad \hatk =\frac{{\bm k}}{\vert{\bm k}\vert}. 
\end{equation}  
The vector functions ${\bm B}_{1,2}(\bdsb{\hat p},\bdsb{\hat k})$ and the tensor function $C^{ij}(\bdsb{\hat p},\bdsb{\hat k})$ can be further decomposed with rotation covariance, with constraints from eq.(\ref{eq:cp2sfi}) under P and CP conservation. Details about this decomposition and constraints can be found in\,\cite{He:2022jjc}.

The general form of the decay amplitude $J/\psi\to B\bar B$ can be written with Lorentz invariance as:
\begin{widetext}
\eqal{
\mathcal{A}^\mu_{J/\psi\to B\bar B} = \bar u(k_1) \left[ \gamma^\mu F_V^B(q^2) + \frac{i}{2m_\Lambda} \sigma^{\mu\nu}q_\nu H_\sigma^B(q^2) + \gamma^\mu\gamma_5 F_A^B(q^2) + \sigma^{\mu\nu}q_\nu \gamma_5 H_T^B(q^2) \right] v(k_2),\label{eq:jpsidecayff}
}
\end{widetext}
with $q=k_1+k_2$ the momentum transfer, and form factors $F_A^B$ and $H_T^B$ the P and the CP violating strength, respectively. $H_\sigma^B$ is the magnetic moment of baryon $B$, whose value is known for all the lowest-lying baryons\,\cite{ParticleDataGroup:2022pth}. Practically, we find the following combinations convenient for expressing the differential angular distributions:
\eqal{
G_1^B = F_V^B + H_\sigma^B, \quad G_2^B = G_1^B - \frac{(k_1 - k_2)^2}{4m_B^2}H_\sigma^B,
}
which are related to the electric and magnetic form factors $G_{E,M}$ and those defined in\,\cite{Faldt:2017kgy} as $G_1 = G_M = {G_M^\psi}/{e_g}$ and $G_2 = G_E = {G_E^\psi}/{e_g}$. To simplify the notations in the subsequent discussion, we leave out the superscript $B$ in the form factors from here on. These form factors $G_{1,2}$ can be determined from the decay parameters of $J/\psi$ as will be explained in section\,\ref{subsec:g12det}.
 
The information of the spin of $B$ or $\bar B$ is revealed in their corresponding  two-body weak decays, 
\textit{i.e.}, $B\to b+\pi$ and its charge conjugated decay $\bar B\to \bar b +\bar\pi$, where $b$ stands for a baryon. For instance, $b$ is the proton for $B=\Lambda$ and $\Lambda$ for $B=\Xi$. The differential decay rates take the following forms\,\cite{Lee:1957qs}:
\eqsp{
   \frac{ d\Gamma_{B}} { d\Omega_b } ({\bm s}_1, \hatl_b )  &\, \propto 1 + \alpha_B  {\bm  s}_1 \cdot \hatl_b,\\ 
   \frac{ d\Gamma_{\bar B}} {d\Omega_{\bar b} } ({\bm s}_2, \hatl_{\bar b} )  &\, \propto 1 -\bar \alpha_B {\bm s}_2 \cdot \hatl_{\bar b}, 
\label{eq:diffdw}   
} 
where $\hatl_b$ and $\hatl_{\bar b}$ are the momentum directions of the secondary baryon $b$ and anti-baryon $\bar b$ in the rest frames of $B$ and $\bar B$,  respectively. $\Omega_{b,\bar b}$ are the corresponding solid angles. $\alpha_B$ and $\bar \alpha_B$ are constants, which satisfies $\alpha_B=\bar\alpha_B$ in the CP conserving limit.

From the differential decay width and $R$ in Eq.(\ref{RJ}), the angular distribution of $e^-e^+ \to J/\psi \to B\bar B$ followed by $B \to b\pi $ and $\bar B\to \bar b \pi$ is given by:
\eqal{
\frac{d\sigma}{ d\Omega_{k} d\Omega_b d\Omega_{\bar b}} \propto  \mathcal{W}({\hatp}, {\hatk}, \hat l_b, \hat l_{\bar b}),\label{eq:diffxsec}
}
where the general from of the angular distribution is given by 
\eqal{
\mathcal{W} = &\, a(\hatp\cdot\hatk) + \alpha_B \hatl_b \cdot \bdsb{B_1}(\bdsb{\hat p},\bdsb{\hat k}) - \bar\alpha_B \hatl_{\bar b}  \cdot \bdsb{B_2}(\bdsb{\hat p},\bdsb{\hat k})\nb\\
&\, - \alpha_B \bar\alpha_B \hat l_b^i \hat l_{\bar b}^j C^{ij}(\bdsb{\hat p},\bdsb{\hat k}),
} 
and $\Omega_k$ is the solid angle of ${\bm k}$. The proportional factors in Eq.(\ref{eq:diffdw}-\ref{eq:diffxsec}) are irrelevant in this work as they cancel in observables, which we will discuss in the following.

In this work we will use integrated observables to study P and CP violation. The four unit vectors $\hatp$, $\hatk$ and $\hatl_{b, \bar b}$ can be used to construct any observable ${\mathcal O}$ for our purpose, 
and its expectation value is given by:  
\eqal{
\langle {\mathcal O}  \rangle = \frac{1}{{\mathcal N}} \int \frac { d\Omega_{k} d\Omega_b d\Omega_{\bar b}} {(4\pi)^3} {\mathcal 
O} \cdot   \mathcal{W}({\hatp}, {\hatk}, \hat l_b, \hat l_{\bar b}),
}
where ${\mathcal N}$ is a normalization factor determined by $\langle 1\rangle =1$. The following expectation values of observables can be used for extracting the spin of $B$ or $\bar B$:
\bewt{
\eqal{
\langle \hatl_{b} \cdot \hatp \rangle &\, = \frac{4\alpha_B}{9\mathcal{N}}E_c^2 d_J \left( {2}y_m {\rm Re}\left(G_1G_2^\ast\right) + \left\vert G_1 \right\vert^2 \right),\label{eq:lpdotp}\\
\langle \hatl_{\bar b} \cdot \hatp \rangle &\, = -\frac{4\bar\alpha_B}{9\mathcal{N}}E_c^2 d_J \left( {2}y_m {\rm Re}\left(G_1G_2^\ast\right) + \left\vert G_1 \right\vert^2 \right),\label{eq:lpbdotp}\\
\langle \hatl_{b} \cdot \hatk\rangle &\,= \frac{8\alpha_B\beta}{9\mathcal{N}}E_c^2 \left[ {\rm Re}\left(F_AG_1^\ast\right) - E_c {\rm Im}\left( y_m H_T G_2^\ast \right) \right],\label{eq:lpdotk}\\
\langle \hatl_{\bar b} \cdot \hatk \rangle  &\,= -\frac{8\bar\alpha_B\beta}{9\mathcal{N}}E_c^2 \left[ {\rm Re}\left(F_AG_1^\ast\right) + E_c {\rm Im}\left( y_m H_T G_2^\ast \right) \right],\label{eq:lpbdotk} \\
 \langle (\hatl_{b} \times \hat l_{\bar b}  ) \cdot \hatk \rangle  &\,= -\frac{16\alpha_B\bar\alpha_B}{27\mathcal{N}}\beta y_m E_c^3{\rm Re}\left( H_TG_2^\ast \right)\label{eq:todd},
}}
with $\beta = \sqrt{1-y_m^2}$ and $y_m = m_B/E_c$. $d_J$ is a measure of P violation on the production side of $J/\psi$, as explained below in eq.\,\eqref{eq:ee2cc}, whose form will be derived in the subsequent section. The normalization factor is given by:
\eqal{
\mathcal{N} = \frac23E_c^2\left( 2\left\vert G_1 \right\vert^2 + y_m^2 \left\vert G_2 \right\vert^2 \right).
}
We comment that there are some errors in the results of the above expectation values in\,\cite{He:2022jjc}. They are corrected here {and given in appendix\,\ref{app:spindecom} for the coefficients ${\bm B}_{1,2}$ and $C^{ij}$ in eq.\,\eqref{eq:spindocom}}. Note also that all expectation values in Eq.(\ref{eq:lpdotp}-\ref{eq:todd}) would vanish if $\alpha_B$ and $\bar \alpha_B$ are zero, \textit{i.e.}, the produced baryon $B$ and $\bar B$ are unpolarized.

Asymmetries can be obtained by the above observables for probing P or CP violating effects.  For any ${\mathcal O}$ one can always define corresponding  asymmetries in events as:
\iffalse
\begin{widetext}
\eqal{
 {\mathcal A}({\mathcal O}  ) = \frac{ {\mathcal N}_{\rm event}  ({\mathcal O}>0) - {\mathcal N}_{\rm event} ({\mathcal O} < 0) } 
    { {\mathcal N}_{\rm event} ({\mathcal O} >0) +  {\mathcal N}_{\rm event} ({\mathcal O} < 0) } = \frac{1}{{\mathcal N}}\int\frac{ d\Omega_k d\Omega_p d\Omega_{\bar p}} { (4\pi)^3}
     \biggr ( \theta ({\mathcal O})- \theta (-{\mathcal O}) \biggr )  {\mathcal W} (\Omega) .  
}   
\end{widetext}
\fi
\begin{widetext}
\eqal{
 {\mathcal A}({\mathcal O}  ) \yong{\equiv \frac{ {\mathcal A}  ({\mathcal O}>0) - {\mathcal A} ({\mathcal O} < 0) } 
    { {\mathcal A} ({\mathcal O} >0) +  {\mathcal A} ({\mathcal O} < 0) }} = \frac{1}{{\mathcal N}}\int\frac{ d\Omega_k d\Omega_p d\Omega_{\bar p}} { (4\pi)^3}
     \biggr ( \theta ({\mathcal O})- \theta (-{\mathcal O}) \biggr )  {\mathcal W} (\Omega),
}   
\end{widetext}
\yong{with $\theta$ the Heaviside function}. The statistical error of any asymmetry \yong{is proportional to} $1/\sqrt{ N_{event}}$ with $N_{event}$ the number of events for the $B\bar B$ final state. {In a real experiment,  the total error will be contributed from several errors besides of the statistical one, e.g., systematic error, momentum resolution, etc. The total one will be only known after 
a detailed Monte Carlo simulation. In our analysis, we will only use the statistical error as a reference value 
for the total one. In the estimation of  $N_{event}$ and the statistical error we take the detector efficiency $\epsilon$ into account.     }
We will focus on the following asymmetries
\eqsp{
\mathcal{A}(\hatl_{b} \cdot \hatp - \hatl_{\bar b} \cdot \hatp) \equiv &\, A_{\rm PV}^{(1)}\\
\simeq &\, \frac{4\alpha_B}{3\mathcal{N}}E_c^2 d_J \left( {2}y_m {\rm Re}\left(G_1G_2^\ast\right) + \left\vert G_1 \right\vert^2 \right),\\
\mathcal{A}( \hatl_{b} \cdot\hatk  - \hatl_{\bar b} \cdot \hatk ) \equiv &\, A_{\rm PV}^{(2)}\\
\simeq &\, \frac{8 \alpha_B\beta}{3\mathcal{N}}E_c^2 \, {\rm Re}\left(F_AG_1^\ast\right),\label{eq:psym}
}
for testing the P symmetry on the production and the decay sides of $J/\psi$, respectively, and the following asymmetries 
\eqsp{
\mathcal{A}( \hatl_{b} \cdot \hatk + \hatl_{\bar b} \cdot \hatk ) &\, \equiv A_{\rm CPV}^{(1)}\simeq -\frac{8\alpha_B\beta}{3\mathcal{N}}E_c^3\, {\rm Im}\left( y_m H_T G_2^\ast \right), \\
\mathcal{A}( ( \hatl_{b} \times \hatl_{\bar b}) \cdot\hatk ) &\, \equiv A_{\rm CPV}^{(2)} \simeq -\frac{8\alpha_B\bar\alpha_B}{9\mathcal{N}}\beta y_m E_c^3{\rm Re}\left( H_TG_2^\ast \right),\label{eq:cpsym}
}
for testing the CP symmetry. We comment that, given the opposite signs in eqs.\,(\ref{eq:lpdotp}-\ref{eq:lpbdotp}), we introduce the difference, \text{i.e.}, $A_{\rm PV}^{(1)}$, as a practical measurement of P violation to enhance its magnitude by a factor of 2. Experimentally, this will require the reconstruction of both $\bdsb{\hat l_{b}}$ and $\bdsb{\hat l_{\bar b}}$. This reconstruction, however, can be performed individually instead of simultaneously such that this enhancement can be achieved without introducing any extra technical difficulties. This is similarly true for $A_{\rm PV}^{(2)}$ and $A_{\rm CPV}^{(1)}$, with the exception for $A_{\rm CPV}^{(2)}$ that requires a simultaneous determination of $\bdsb{\hat l_{b}}$ and $\bdsb{\hat l_{\bar b}}$. Interestingly, one also finds that $A_{\rm CPV}^{(1)}$ and $A_{\rm CPV}^{(2)}$ individually probes the imaginary and the real parts of $H_TG_2^\ast$.

Quantitative predictions for these asymmetries rely on prior knowledge of these form factors and the decay parameters. The latter has been measured, at BESIII for instance, for some of the lowest-lying baryons as we will summarize in section\,\ref{sec:res}, the former needs to be derived theoretically to connect theories with experiments, to which we devote the next section.

%%%%%%%%%%%%%%%%%%%%%
\section{Form factors}\label{sec:formfac}
%%%%%%%%%%%%%%%%%%%%%
The asymmetries in eqs.\,(\ref{eq:psym}-\ref{eq:cpsym}) depend on the form factors $F_A$, $G_{1,2}$ and $H_T$ defined in last section, which play a key role in reinterpreting data in theories. The form factors describing $J/\psi\to B\bar B$ at low energies generically receive non-perturbative contributions, and can be calculated with nonperturbative methods. In this section, we give a derivation for these form factors from a more phenomenological approach.

\begin{table*}[!htb]\caption{Contributions to $d_B$ from quark EDMs, \textit{i.e.}, $d_q$, in the quark model (second column) and quark CDMs, \textit{i.e.}, $f_q$, in the non-relativistic (NR) quantum chromodynamics (QCD) limit in the quark model (fifth column). Given the current constraints on $d_{n,p}$ as detailed in the main text, the reduced results are given, respectively, in the third and the sixth columns.}\label{tab:qedm2dB}
\centering
\begin{ruledtabular}
\begin{tabular}{ccc|ccc}
$d_B$   & QM & Reduced Results & $d_B$   & NR QCD \& QM & Reduced Results \\
\hline
$d_p^{\rm qEDM}$     & $\frac{1}{3}(4d_u - d_d)$  & --- & $d_p^{\rm\,qCDM}$ & $-\frac{1}{3} \left( 4Q_d f_d - Q_u f_u \right)$ & --- \\
$d_n^{\rm qEDM}$     & $\frac{1}{3}(4d_d - d_u)$  & --- & $d_p^{\rm\,qCDM}$ &  $-\frac{1}{3} \left( 4Q_u f_u - Q_d f_d \right)$ & --- \\
$d_{\Sigma^+}^{\rm qEDM}$     & $\frac{1}{3}(4d_u - d_s)$  & $ -\frac13d_s$ & $d_p^{\rm\,qCDM}$ &  $-\frac{1}{3} \left( 4Q_u f_u - Q_s f_s \right)$ & $-\frac19ef_s$ \\
$d_{\Sigma^0}^{\rm qEDM}$     & $\frac{1}{3}(2d_u + 2d_d - d_s)$  & $ -\frac13d_s$ & $d_p^{\rm\,qCDM}$ &  $-\frac{1}{3} \left( 2Q_u f_u + 2Q_d f_d - Q_s f_s \right)$ & $-\frac19ef_s$ \\
$d_{\Sigma^-}^{\rm qEDM}$     & $\frac{1}{3}(4d_d - d_s)$  & $ -\frac13d_s$ & $d_p^{\rm\,qCDM}$ &  $-\frac{1}{3} \left( 4Q_d f_d - Q_s f_s \right)$ & $-\frac19ef_s$ \\
$d_{\Xi^0}^{\rm qEDM}$     & $\frac{1}{3}(4d_s - d_u)$  & $\frac43d_s$ & $d_p^{\rm\,qCDM}$ &  $-\frac{1}{3} \left( 4Q_s f_s - Q_u f_u \right)$ & $\frac49ef_s$ \\
$d_{\Xi^-}^{\rm qEDM}$     & $\frac{1}{3}(4d_s - d_d)$  & $\frac43d_s$ & $d_p^{\rm\,qCDM}$ &  $-\frac{1}{3} \left( 4Q_s f_s - Q_d f_d \right)$ & $\frac49ef_s$ \\
$d_{\Lambda^0}^{\rm qEDM}$     & $d_s$  & $d_s$ & $d_p^{\rm\,qCDM}$ &  $-Q_s f_s$ & $\frac13ef_s$ \\
\end{tabular}
\end{ruledtabular}
\end{table*}
%%%%%%%%
\subsection{{$H_T$ from electric dipole interactions}}\label{subsec:htres}
%%%%%%%%
The form factor $H_T$ depends on the momentum transfer $q^2$ and is related to the electric dipole moment (EDM) of $B$ in the vanishing $q^2$ limit. For this work, we ignore the $q^2$ dependence of $H_T$ and approximate it as the real EDM of baryon $B$ in the final state of the decaying $J/\psi$, \textit{i.e.},
\eqal{
\mathcal{L}_{B_{\rm EDM}} = -i\frac {d_B}{2}\bar B \sigma_{\mu\nu}\gamma_5 B F^{\mu\nu},\label{eq:lambdaedm}
}
with $d_B$ the baryon EDM, $F^{\mu\nu}$ the electromagnetic field tensor, and $B$ the baryon field. In this approximation, the decay amplitude of $J/\psi\to B\bar B$ is given by
\eqal{
i\mathcal{M}_{\rm decay} &\, = \frac{ieQ_c d_B}{m_{J/\psi}^2} \cdot \left(\bar B\sigma^{\alpha\beta} q_\beta \gamma_5 B \right) \cdot g_V \cdot \epsilon_\alpha^{J/\psi},\label{eq:edm2jpsi}
}
where $Q_c=2/3$ is the electric charge of the $c$ quark, and we use $\bra{0}(\bar c\gamma^\alpha c ) \ket{J/\psi}=g_V \cdot \epsilon_\alpha^{J/\psi}$ for the charm quark vector current with the form factor $g_V=\left(1.286\pm0.012\right){\rm\,GeV^2}$ as determined from the branching ratio of $J/\psi\to \ell^\pm\ell^\mp$\,\cite{He:1992ng}:
\eqal{
\Gamma_{J/\psi\to\ell^\pm\ell^\mp} = \frac{16g_V^2\pi \alpha_{\rm EM}^2 }{27m_{J/\psi}^3}\left( 1 + \frac{2 m_{\ell}^2}{m_{J/\psi}^2} \right)\sqrt{1 - \frac{4 m_{\ell}^2}{m_{J/\psi}^2}},\label{eq:jpsilep}
}
with $\alpha_{\rm EM}=e^2/4\pi$ the fine structure constant and $e$ the electron charge magnitude. From eq.\,\eqref{eq:edm2jpsi}, its matching with eq.\,\eqref{eq:jpsidecayff} leads to
\eqal{
H_T = \frac{e \cdot Q_c \cdot g_V \cdot d_B}{m_{J/\psi}^2}.
}
To connect with the quark-level interactions, we further estimate the baryon EDM $d_B$ by decomposing it into contributions from its quark components in the quark model (QM). In this respect, we consider two different sources: (1) EDMs of quarks ($d_q$), and (2) chromo-dipole moments (CDMs) of quarks ($f_q$). The results are summarized in the second and fifth columns of table\,\ref{tab:qedm2dB}, whose derivation is given in appendix\,\ref{app:dBres}.

Recall that the current upper bounds on $d_n$, $d_p$, and $d_\Lambda$ are, respectively, $|d_n|<1.8\times10^{-26}\,e\,{\rm cm}$\,\cite{Abel:2020pzs}, $|d_p|<2.1\times10^{-25}\,e\,{\rm cm}$\,\cite{ParticleDataGroup:2022pth}, and $|d_{\Lambda}| < 1.5\times10^{-16}\,e\,{\rm cm}$\,\cite{Pondrom:1981gu}, the different combinations of $d_{n,p}$ in terms of $d_{u,d}$ or $f_{u,d}$ would suggest the dominance of the strange quark in contributing to the baryon EDMs in the quark model. Upon this observation, one approximately finds the following ``reduced'' relations
\eqal{
&\, d_{\Sigma^{0,\pm}}^{\rm qEDM,\,qCDM} \simeq  -\frac13d_{\Lambda^0}^{\rm qEDM,\,qCDM},\nb\\
&\, d_{\Xi^{0,-}}^{\rm qEDM,\,qCDM} \simeq \frac43d_{\Lambda^0}^{\rm qEDM,\,qCDM},\label{eq:baredm2qedm}
}
as shown in the third and sixth columns of table\,\ref{tab:qedm2dB}. Based on these reduced results, the 95\% confidence level (CL) upper bounds on $d_B$ are thus estimated to be of $\mathcal{O}(10^{-17}\sim10^{-16})\,e{\rm\,cm}$ based on that of $d_\Lambda$ in\,\cite{Pondrom:1981gu}. As we will see shortly in section\,\ref{subsec:psym}, these upper bounds are already expected to be improved at BESIII or STCFs utilizing $A_{\rm CPV}^{(1,\,2)}$ introduced in this work.

\begin{figure*}
\begin{center}
\begin{minipage}{\textwidth}
\includegraphics[scale=0.5]{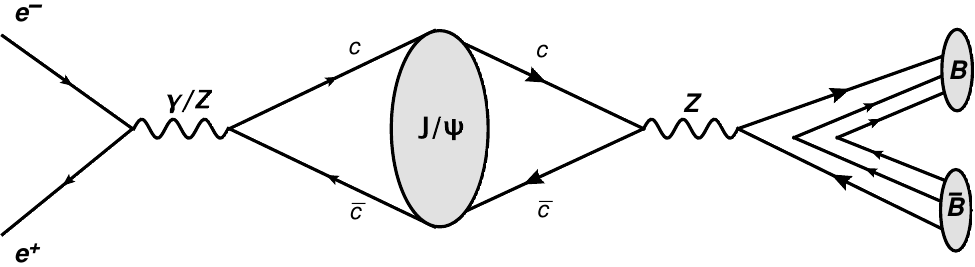} \qquad\qquad \includegraphics[scale=0.44]{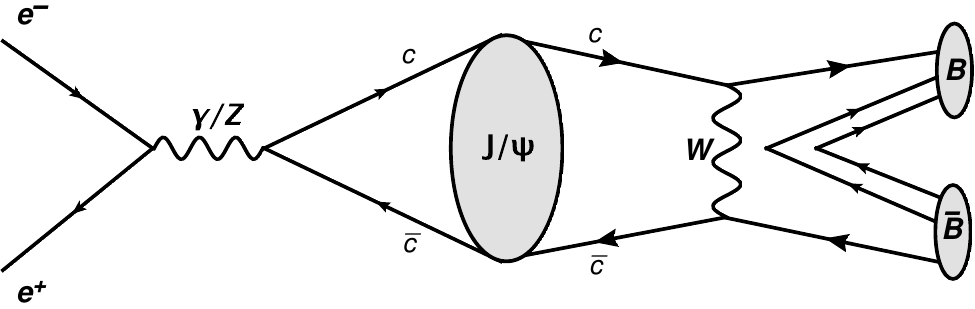} \\
\caption{{Schematic diagrams for $e^+e^-\to J/\psi\to B\bar B$ via an intermediate $Z$ (left) and an intermediate $W$ (right). The gray blobs represent bound states as given at the centers.}}\label{fig:feyndia}
\end{minipage}
\end{center}
\end{figure*}

%%%%%%%%
\subsection{{$F_A$ from P violating interactions}}
%%%%%%%%
The form factor $F_A$ originates from the parity-violating decay of $J/\psi$, which is rooted in an intermediate $s$-channel $Z$ and an intermediate $t$-channel $W$ boson exchanges in the SM. {The schematic diagrams are explicitly shown in the left and right panels of Fig.\,\ref{fig:feyndia}, respectively, for clarification}. The quark level Lagrangian in the low momentum transfer limit, as is the case for $J/\psi$ production and decay considered in this work, is given by
\begin{widetext}
\eqal{
\mathcal{L}_{Z+W} = &\, -4\sqrt2 G_F \cdot \sum_{q=u,d,s}\left[ g_{V-A}^qg_{V-A}^c \left( \bar q_R\gamma_\mu q_R \right)\left( \bar c_R\gamma_\mu c_R \right) + g_{V-A}^qg_{V+A}^c \left( \bar q_R\gamma_\mu q_R \right) \left( \bar c_L\gamma_\mu c_L \right)\right.\nb\\
&\, \left. \qquad\qquad\quad + g_{V+A}^qg_{V-A}^c \left( \bar q_L\gamma_\mu q_L \right)\left( \bar c_R\gamma_\mu c_R \right) + g_{V+A}^qg_{V+A}^c \left( \bar q_L\gamma_\mu q_L \right) \left( \bar c_L\gamma_\mu c_L \right) \right]\nb\\
&\, - 2\sqrt2 G_F \left[ \left\vert V_{cd} \right\vert^2 \left(\bar d_L \gamma_\alpha c_L \right) \left(\bar c_L \gamma^\alpha d_L\right) + \left\vert V_{cs} \right\vert^2 \left( \bar s_L \gamma_\alpha c_L \right) \left(\bar c_L \gamma^\alpha s_L\right)  \right],\label{eq:4fwz}
}
\end{widetext}
where the first term is from an intermediate $Z$ exchange and the second that from an intermediate $W$. {For the latter case, the bottom quark is excluded from the sum since the corresponding amplitude is suppressed by $|V_{cb}|^2\approx10^{-3}$\,\cite{ParticleDataGroup:2022pth}.} $G_F$ is the Fermi constant as determined from the muon lifetime, $V$ is the CKM matrix, {$f_{L\,(R)}\equiv P_{L\,(R)}f$ are the chiral fields with $P_{L,\,(R)}$ the chiral projectors}, and the couplings are all defined at the matching scale $m_Z$ as given by
\eqal{
g_{V\pm A}^q \equiv g_{V}^q \pm g_{A}^q, \quad g_V^q = \frac12 T_q^3 - Q_q s_w^2, \quad g_A^q = \frac12T_q^3.
}
Here, $c_w\equiv\cos\theta_w$ and $s_w\equiv\sin\theta_w$ with $\theta_w$ the weak mixing angle, $T_q^3$ and $Q_q$ are, respectively, the iso-spin and the electric charge of quark $q$. Since $J/\psi\to B\bar B$ occurs at a much lower energy scale than $m_Z$, the running effects on these operators will have to be included due to significant QCD modifications. These have been well documented in literature, see for example, Refs.\,\cite{Gaillard:1974nj,Gaillard:1974nj,Ecker:1985vv,He:1988th,Buchalla:1995vs} for kaon decay and ref.\,\cite{Jenkins:2017dyc} for the weak effective field theory. For this work, we extract the anomalous dimension from\,\cite{Jenkins:2017dyc} by focusing on the dominant pure QCD corrections only. Since the singlet operators in eq\,\eqref{eq:4fwz} will mix with the octet ones through the anomalous dimension even under this approximation, we define the following linear combinations to decouple them such that the anomalous dimension in this basis is diagonal:
\bewt{
\eqsp{
% w case
&\, \mathcal{O}_{\pm}^{LL\,(RR)} = \left(\bar d \gamma_\mu P_{L\,(R)} c\right)\left(\bar c \gamma_\mu P_{L\,(R)} d\right) \pm \left(\bar d \gamma_\mu P_{L\,(R)} d\right)\left(\bar c \gamma_\mu P_{L\,(R)} c\right),\\
&\, \mathcal{O}_{+}^{LR\,(RL)} = \left(\bar d \gamma_\mu P_{L\,(R)} c\right)\left(\bar c \gamma_\mu P_{R\,(L)} d\right) + \frac23 \left(\bar d P_{R\,(L)} d\right)\left(\bar c P_{L\,(R)} c\right),\\
&\, \mathcal{O}_{-}^{LR\,(RL)} = \frac23 \left(\bar d P_{R\,(L)} d\right)\left(\bar c P_{L\,(R)} c\right),
}
}
for $W$, and 
\bewt{
{\allowdisplaybreaks
\begin{align*}
% z case
&\, \mathcal{O}_{uu+}^{LR\,(RL)} = \frac{1}{N_c} \left( \bar u_{R\,(L)}\gamma_\mu u_{R\,(L)} \right) \left( \bar c_{L\,(R)}\gamma_\mu c_{L\,(R)} \right) + 2 \left( \bar u_{R\,(L)}\gamma_\mu T^A u_{R\,(L)} \right) \left( \bar c_{L\,(R)}\gamma_\mu T^A c_{L\,(R)} \right),\\
&\, \mathcal{O}_{uu-}^{LR\,(RL)} = -\frac{4 C_F}{N_c} \left( \bar u_{R\,(L)}\gamma_\mu u_{R\,(L)} \right) \left( \bar c_{L\,(R)}\gamma_\mu c_{L\,(R)} \right) + \frac{4}{N_c} \left( \bar u_{R\,(L)}\gamma_\mu T^A u_{R\,(L)} \right) \left( \bar c_{L\,(R)}\gamma_\mu T^A c_{L\,(R)} \right),\\
&\, \mathcal{O}_{ud+}^{LR} = \frac{1}{N_c} \left( \bar d_R\gamma_\mu d_R \right) \left( \bar c_L\gamma_\mu c_L \right) + 2 \left( \bar d_R\gamma_\mu T^A d_R \right) \left( \bar c_L\gamma_\mu T^A c_L \right),\stepcounter{equation}\tag{\theequation}\\
&\, \mathcal{O}_{ud-}^{LR} = -\frac{4 C_F}{N_c} \left( \bar d_R\gamma_\mu d_R \right) \left( \bar c_L\gamma_\mu c_L \right) + \frac{4}{N_c} \left( \bar d_R\gamma_\mu T^A d_R \right) \left( \bar c_L\gamma_\mu T^A c_L \right),\\
&\, \mathcal{O}_{du+}^{LR} = \frac{1}{N_c} \left( \bar d_L\gamma_\mu d_L \right) \left( \bar c_R\gamma_\mu c_R \right) + 2 \left( \bar d_L\gamma_\mu T^A d_L \right) \left( \bar c_R\gamma_\mu T^A c_R \right),\\
&\, \mathcal{O}_{du-}^{LR} = -\frac{4 C_F}{N_c} \left( \bar d_L\gamma_\mu d_L \right) \left( \bar c_R\gamma_\mu c_R \right) + \frac{4}{N_c} \left( \bar d_L\gamma_\mu T^A d_L \right) \left( \bar c_R\gamma_\mu T^A c_R \right),
\end{align*}
}}for $Z$, with $T^A$ the Gell-Mann matrices normalized as ${\rm Tr}(T^A T^B) = \delta^{AB}/2$, $C_F = (N_c^2-1)/(2N_c)$, $N_c=3$ the number of colors. 

Using these combinations, we rewrite eq.\,\eqref{eq:4fwz} as
\bewt{
\eqsp{
-\mathcal{L}_{Z+W} = &\, 4\sqrt2 G_F \cdot \sum_{q=u,d,s}\left[ \frac12 g_{V-A}^qg_{V-A}^c \left( C_+^R \mathcal{O}_{+}^{RR} - C_-^R \mathcal{O}_{-}^{RR}\right) + \frac12 g_{V+A}^qg_{V+A}^c \left( C_+^L \mathcal{O}_{+}^{LL} - C_-^L \mathcal{O}_{-}^{LL}\right)\right]\\
&\, + 4\sqrt2 G_F \cdot \left[ \frac{g_{V-A}^u g_{V+A}^c}{N_c} C_{uu+}^{LR} \mathcal{O}_{uu+}^{LR} - \frac{g_{V-A}^u g_{V+A}^c }{2} C_{uu-}^{LR} \mathcal{O}_{uu-}^{LR}\right.\\
&\, \left.\qquad\qquad\qquad + \frac{g_{V+A}^u g_{V-A}^c}{N_c} C_{uu+}^{RL} \mathcal{O}_{uu+}^{RL} - \frac{g_{V+A}^u g_{V-A}^c}{2} C_{uu-}^{RL} \mathcal{O}_{uu-}^{RL} \right]\\
&\, + 4\sqrt2 G_F \cdot \sum_{q=d,s}\left[ \frac{g_{V-A}^q g_{V+A}^c}{N_c} C_{ud+}^{LR} \mathcal{O}_{ud+}^{LR} - \frac{g_{V-A}^q g_{V+A}^c }{2} C_{ud-}^{LR} \mathcal{O}_{ud-}^{LR}\right.\\
&\, \left.\qquad\qquad\qquad + \frac{g_{V+A}^q g_{V-A}^c}{N_c} C_{du+}^{LR} \mathcal{O}_{du+}^{LR} - \frac{g_{V+A}^q g_{V-A}^c}{2} C_{du-}^{LR} \mathcal{O}_{du-}^{LR} \right]\\
&\, + \sqrt2 G_F \sum_{q=d,s} \left\vert V_{cq}\right\vert^2 \left( C_+^L \mathcal{O}_{+}^{LL} + C_-^L \mathcal{O}_{-}^{LL}\right),\label{eq:4fwz2}
}}
with all the $C$'s being unity at scale $m_Z$. Running to an arbitrary scale $\mu$, we find
\bewt{
\eqsp{
&\, C_+^{L,R}(\mu^2) \equiv C_1\left(\mu^2\right) = \left( \frac{\alpha_s\left(\mu^2\right)}{\alpha_s\left(m_Z^2\right)} \right)^{-\frac{3(N_c-1)}{b N_c}},\quad C_-^{L,R}(\mu^2) \equiv C_2\left(\mu^2\right) = \left( \frac{\alpha_s\left(\mu^2\right)}{\alpha_s\left(m_Z^2\right)} \right)^{\frac{3(1+N_c)}{b N_c}},\\
&\, C_{uu+,ud+,du+}^{LR}\left(\mu^2\right) \equiv C_3\left(\mu^2\right) = \left( \frac{\alpha_s\left(\mu^2\right)}{\alpha_s\left(m_Z^2\right)} \right)^{\frac{6C_F}{b}},\quad C_{uu-,ud-,du-}^{LR}\left(\mu^2\right) \equiv C_4\left(\mu^2\right) = \left( \frac{\alpha_s\left(\mu^2\right)}{\alpha_s\left(m_Z^2\right)} \right)^{-\frac{3}{bN_c}},
}}
with the strong coupling $\alpha_s$ and its $\beta$ function $b$ given by
\eqal{
\alpha_s(\mu^2) = \frac{4\pi}{b\log\left(\frac{\mu^2}{\Lambda_f^2}\right)}, \quad b = \frac{11}{3}N_c - \frac23N_f,
}
where $\Lambda_f$ is the QCD parameter for $N_f$ dynamical quark flavors under consideration in the $\overline{\rm MS}$ scheme, and $\mu$ is the 't Hooft scale below $m_Z$. For $J/\psi$ decay as considered in this work, since $m_{J/\psi}\approx m_b$ with $m_b$ the bottom quark mass, we consider the running of these parameters down to the scale $\mu = m_{J/\psi}$ with $N_f = 5$, $\Lambda_5=225\rm\,MeV$\,\cite{Buchalla:1995vs}, and we use the average $\alpha_s(m_Z^2) =  0.1180$ as reported in\,\cite{ParticleDataGroup:2022pth}. Numerically, we find
\eqsp{
&\, C_1(m_{J/\psi}^2) = 0.78, \quad C_2(m_{J/\psi}^2) = 1.66,\\
&\, C_3(m_{J/\psi}^2) = 2.76, \quad C_4(m_{J/\psi}^2) = 0.88\,,\label{eq:cvals}
}
which are significantly modified compared with their values at $\mu=m_Z$. {On the other hand, since $m_b$ is slightly above $m_{J/\psi}$, integrating also the $b$ quark out leads to a slightly different setup with
\eqsp{
&\, C_1(m_{J/\psi}^2) = 0.78,\quad C_2(m_{J/\psi}^2) = 1.65,\\
&\, C_3(m_{J/\psi}^2) = 2.72, \quad C_4(m_{J/\psi}^2) = 0.88,
}
with $N_f = 4$ and $\Lambda_4=325\rm\,MeV$\,\cite{Buchalla:1995vs}. The axial-vector form factors with this setup, which will be discussed shortly below in eq.\,\eqref{eq:farge}, are found mildly modified, and our main conclusion from this study is thus barely affected.} In the following, to simplify the notations, we rename
\eqsp{
C_1(m_{J/\psi}^2) = c_1,\quad C_2(m_{J/\psi}^2) = c_2,\\
C_3(m_{J/\psi}^2) = c_3, \quad C_4(m_{J/\psi}^2) = c_4,\label{eq:cirge}
}
{and use their values in eq.\,\eqref{eq:cvals} for the following discussion.} Then by including the running effects and focusing on the parity violating part of the decay amplitude for the extraction of $F_A$, we obtain {with vacuum saturation\,\cite{Donoghue:2022wrw}}
\bewt{
\eqsp{
i\mathcal{M}_{F_A} = &\, 4\sqrt2G_F \cdot i g_V \cdot \epsilon_\alpha^{J/\psi} \\
&\, \times {\bra{B\bar B} \left[ \frac12\sum_{q=u,d,s} \left( g_V^cg_A^q + g_A^c g_V^q \right)\mathcal{R}_Z + \frac12\sum_{q=u,d,s} \left( g_V^cg_A^q - g_A^c g_V^q \right)\mathcal{\tilde R}_Z + \sum_{q=d,s} \frac{\left\vert V_{cq}\right\vert^2}{8} \mathcal{R}_W \right] \bar q \gamma^\alpha \gamma_5 q \ket{0}}, \label{eq:faextract}
}}
with 
\eqsp{
&\, \mathcal{R}_Z = \frac{1}{2}\left[ (c_1 + c_2) + \frac{c_1 - c_2}{N_c}\right] \approx 1.07,\\
&\, \mathcal{\tilde R}_Z = c_4 + \frac{c_3-c_4}{N_c^2} \approx 1.09,\\
&\, \mathcal{R}_W = \frac{1}{2}\left[ (c_1 - c_2) + \frac{c_1 + c_2}{N_c}\right] \approx -0.03,\label{eq:rgecorfac}
}
corresponding to about 7\%, 9\% enhancement and a factor of 10 reduction, respectively, compared with their values at $m_Z$. It is straightforward to check that one recovers the quark level Lagrangian in eq.\,\eqref{eq:4fwz} by taking the unity limits of $c_i$'s at $\mu=m_Z$ in eq.\,\eqref{eq:faextract}, which also provides a cross check of our results. We comment that contributions from $\mathcal{O}_{uu\pm}^{LR\,(RL)}$ completely cancel out due to lepton flavor universality, we nevertheless include them in eq.\,\eqref{eq:faextract} for compactness and a straightforward generalization to new physics analysis where lepton flavor universality is not necessarily respected. It is also noted that compared with the scenario without QCD running where only $g_V^c g_A^q$'s (with $q=u,d,s$) contribute to $F_A$ in the $Z$ process, the axial-vector current of the $c$ quark also gives a non-vanishing contribution when the renormalization group evolution is included.

The evaluation of $\bra{B\bar B} \left(\bar q \gamma^\alpha \gamma_5 q \right)\ket{0}$ is {estimated} by computing the crossed matrix element $\bra{B} \left(\bar q \gamma^\alpha \gamma_5 q \right)\ket{B}$ using chiral perturbation theory ($\chi$PT). The details can be found in appendix\,\ref{app:fader}, and we finally obtain
\bewt{{\allowdisplaybreaks
\begin{align*}
F_A^{n} &\, = \left(\frac{G_F g_V}{2\sqrt2}\right) \cdot \left\{ D \cdot \left[ \left( 1- 2s_w^2 \right)\mathcal{R}_Z - \frac23 s_w^2 \mathcal{\tilde R}_Z \right]  - \left[ D +  \left( \left\vert V_{cd}\right\vert^2 - \left\vert V_{cs}\right\vert^2 \right) F \right] \mathcal{R}_W \right\},\\
\quad F_A^{p} &\, = - \left(\frac{G_F g_V}{2\sqrt2}\right) \cdot \left\{ \left( F -\frac{D+7F}{3}s_w^2 \right) \mathcal{R}_Z + \frac{D-F}{3} s_w^2 \mathcal{\tilde R}_Z  + (D-F) \left\vert V_{cs}\right\vert^2 \mathcal{R}_W  \right\},\\
F_A^{\Sigma^+} &\, = -\left(\frac{G_F g_V}{2\sqrt2}\right) \cdot \left\{ \left( F -\frac{D+7F}{3}s_w^2 \right) \mathcal{R}_Z + \frac{D-F}{3} s_w^2 \mathcal{\tilde R}_Z  + (D-F) \left\vert V_{cd}\right\vert^2 \mathcal{R}_W \right\},\\
F_A^{\Sigma^0} &\, =  \left(\frac{G_F g_V}{2\sqrt2}\right) \cdot D \cdot \left\{ \frac13s_w^2\left( \mathcal{R}_Z - \mathcal{\tilde R}_Z \right) - \left\vert V_{cd}\right\vert^2 \mathcal{R}_W \right\},\stepcounter{equation}\tag{\theequation}\label{eq:farge}\\
F_A^{\Sigma^-} &\, = \left(\frac{G_F g_V}{2\sqrt2}\right) \cdot \left\{ \left[ \left( \frac{D-7F}{3}s_w^2 + F \right) \mathcal{R}_Z - \frac{D+F}{3} s_w^2 \mathcal{\tilde R}_Z \right]   - (D+F) \left\vert V_{cd}\right\vert^2 \mathcal{R}_W \right\},\\
F_A^{\Xi^0} &\, = \left(\frac{G_F g_V}{2\sqrt2}\right) \cdot \left\{ D \cdot \left[ \left( 1- 2s_w^2 \right)\mathcal{R}_Z - \frac23 s_w^2 \mathcal{\tilde R}_Z \right]  - \left[ D -  \left( \left\vert V_{cd}\right\vert^2 - \left\vert V_{cs}\right\vert^2 \right) F \right] \mathcal{R}_W \right\},\\
F_A^{\Xi^-} &\, = \left(\frac{G_F g_V}{2\sqrt2}\right) \cdot \left\{ \left[ \left( \frac{D-7F}{3}s_w^2 + F \right) \mathcal{R}_Z - \frac{D+F}{3} s_w^2 \mathcal{\tilde R}_Z \right]   - (D+F) \left\vert V_{cs}\right\vert^2 \mathcal{R}_W \right\},\\
F_A^{\Lambda} &\, = \left(\frac{G_F g_V}{3\sqrt2}\right) \cdot D \cdot \left\{ \left[ \left( 1-\frac{11}{6}s_w^2 \right) \mathcal{R}_Z -\frac56 s_w^2\mathcal{\tilde R}_Z \right] - \frac{1}{2} \left( \left\vert V_{cd}\right\vert^2 + 4 \left\vert V_{cs}\right\vert^2 \right) \mathcal{R}_W \right\},
\end{align*}
}}where the unitary condition $\left\vert V_{cd}\right\vert^2 + \left\vert V_{cs}\right\vert^2\approx1$ is applied in obtaining the results above. Note that the contribution to $F_A^{\Sigma^0}$ from the $s$-channel $Z$ exchange from the renormalization group evolution would vanish at $\mu=m_Z$ since at which $\mathcal{R}_Z = \mathcal{\tilde R}_Z = 1$. As a consequence, $F_A^{\Sigma^0}$ would be sensitive to the underlying $Z$ exchange process, especially given the $V_{cd}$ suppression for the $W$ channel.

\begin{table*}[htb]\caption{Numerical results of axial-vector form factors for different baryons, where the first numeric line with label ``No'' indicates results where the running effects discussed in the main text are excluded from the calculations. Corrections to the $t$-channel $W$ process are shown in the ``$t$'' row, and to both the $t$-channel $W$ and the $s$-channel $Z$ processes in the ``$t+s$'' row.}\label{tab:fanumres}
\centering
\begin{ruledtabular}
\begin{tabular}{c|cccccccc}
\backslashbox{Running?}{$F_A^B$}       & $F_A^{n}\,(\times10^{-6})$ & $F_A^{p}\,(\times10^{-7})$ & $F_A^{\Sigma^+}\,(\times10^{-7})$ & $F_A^{\Sigma^0}\,(\times10^{-9})$ & $F_A^{\Sigma^-}\,(\times10^{-7})$ & $F_A^{\Xi^0}\,(\times10^{-6})$ & $F_A^{\Xi^-}\,(\times10^{-7})$ & $F_A^{\Lambda}\,(\times10^{-7})$ \\
\hline
No     & 0.79 & $-11.0$  & $-8.22$ & $-49.9$  & 7.22 & $-0.55$ & $-8.93$ &  $-5.84$ \\
$t$     & 1.18 & $-7.78$ & $-8.06$ & $4.99$   & 8.16 & 1.31       & 9.77      &  8.95      \\
$t+s$ & 1.25 & $-8.39$ & $-8.66$ & $1.25$   & 8.69 & 1.39       & 10.3      &  9.44      \\
\end{tabular}
\end{ruledtabular}
\end{table*}

Numerically, form factors for different baryons are listed in table\,\ref{tab:fanumres}, where the ``No'' row corresponds to results without considering any renormalization group evolution. The ``$t$'' row represents results with the $t$-channel $W$ exchange process improved by the renormalization group equations, and the ``$s+t$'' row that where the additional $s$-channel $Z$ exchange process is also improved. As is clear from the table, the parity-violating form factors $F_A$ from the decay of $J/\psi$ are generically modified by a factor of a few when the running effects are included, dominantly rooted in large QCD corrections for the non-local $t$-channel process. In contrast, the additional $s$-channel corrections turn out to be relatively small, as can be seen from the ``$s+t$'' row of table\,\ref{tab:fanumres}. This observation is true for almost all the octet baryons with the only exception being $\Sigma^0$, and the reason is clear from a closer look at eqs.\,(\ref{eq:rgecorfac}-\ref{eq:farge}): When the running effects are not included, $c_i=1$ ($i=1,\cdots,4$) such that $F_A^{\Sigma^0}\propto \mathcal{R}_W=1/N_c$. In contrast, if only the ``$t$'' process is improved by the renormalization group equations, even though it still hods that $F_A^{\Sigma^0}\propto \mathcal{R}_W$, the accidental cancellation between the first term $(c_1-c_2)$ and the second term $(c_1+c_2)/N_c$ in $\mathcal{R}_W$ of eq.\,\eqref{eq:rgecorfac} reduces its value by a factor of 10 in addition to the sign flip. Furthermore, $F_A^{\Sigma^0}$ is further bated by a factor of $\sim4$ when the renormalization group improved $s$-channel $Z$ exchange process is included, such that its magnitude becomes $\sim10^{3}$ smaller than its charged counterparts, {\textit{i.e.}}, $\Sigma^\pm$. This makes it very challenging to test the P symmetry through $J/\psi\to\Sigma^0\bar\Sigma^0$ unless augmented by new P violating sources.

%%%%%%%%
\subsection{{$G_{1,2}$ from the decay of $J/\psi$}}\label{subsec:g12det}
%%%%%%%%
We discuss the phenomenological determination of the remaining decay form factors $G_{1,2}$ in this section. Due to the smallness of P and P violation, contributions from the $F_A$ and the $H_T$ terms in eq.\,\eqref{eq:jpsidecayff} to the decay width of $J/\psi$ can be safely neglected such that
\eqal{
\Gamma_{J/\psi\to B\bar B} = \frac{|G_1|^2 m_{J/\psi} }{12 \pi}\sqrt{1-\frac{4m_B^2}{m_{J/\psi}^2}}\left( 1+  \frac{2m_B^2}{m_{J/\psi}^2} \left\vert \frac{G_2}{G_1} \right\vert^2 \right).
}
Recall that
\eqal{
\frac{G_2}{G_1} = \frac{G_E^\psi}{G_M^\psi} = \frac{G_E}{G_M} = \left\vert \frac{G_E}{G_M} \right\vert e^{i\Delta\Phi} \equiv e^{i\Delta\Phi} \mathtt{R},\label{eq:rdelphi}
}
with $\Delta\Phi$ the relative phase between $G_{2}$ and $G_1$, and the ratio $\mathtt{R}$ experimentally determined from the angular distribution parameter {$\alpha_{J/\psi}^B$ defined as the ratio of electromagnetic form factors\,\cite{Faldt:2017kgy}}
\eqal{
{\alpha_{J/\psi}^B} = \frac{s \left\vert G_M^\psi \right\vert^2 - 4m_B^2 \left\vert G_E^\psi \right\vert^2}{s \left\vert G_M^\psi \right\vert^2 + 4m_B^2 \left\vert G_E^\psi \right\vert^2} = \frac{s - 4m_B^2\mathtt{R}^2}{s + 4m_B^2 \mathtt{R}^2},
}
where $s=4E_c^2$, one can readily determine $G_1$ from $\Gamma_{J/\psi\to B\bar B}$. Since only the relative phase $\Delta\Phi$ is of interest for this study, without loss of generality, $G_1$ can be taken as a real parameter as will be adopted in the following.

%%%%%%%%
\subsection{{$d_J$ from the production of $J/\psi$}}\label{subsec:djdet}
%%%%%%%%
Finally, we derive the parity violating parameter $d_J$ from the production of on-shell $J/\psi$ particles in this section. We start by parameterizing the production amplitude of $J/\psi$ as
\eqal{
i\mathcal{M} = \frac{ie^2Q_c Q_e}{s}\left[\bar v_{e^+}\left( \gamma_\mu + d_J \gamma_\mu\gamma_5 \right) u_{e^-}\right] \left[ \bar u_{c}\gamma^\mu v_{\bar c} \right],\label{eq:ee2cc}
}
with $d_J$ the parity-violating part of the amplitude. Clearly, the only tree-level contribution to $d_J$ in the SM comes from the $s$-channel $Z$ exchange diagram that violates parity, and a direct matching between the amplitudes leads to
\eqal{
d_J = \frac{\sqrt2 s G_F}{32\pi\alpha_{\rm EM}}\cdot (3 - 8 s_w^2).\label{eq:djres}
}
It shall be understood that $d_J$ above is obtained at scale $m_Z$ and thus shall also be evolved to scale $m_{J/\psi}$ in analogy to our discussion on $F_A$ using the renormalization group equations. Recall that since the quark bilinear is an SU(3)$_c$ singlet, pure QCD corrections at the leading order will be absent in this case. Thus to the leading order, these semi-leptonic operators only receive mixed quantum electrodynamics (QED) and QCD corrections from the dipole operators that arise at the loop order in the SM and are also model-independently very stringently constrained experimentally. We therefore ignore the running effects on $d_J$ and interpret it as the P violating parameter at the desired scale.

With these results, we close the discussion on form factors and further rewrite P and CP violating asymmetries defined above into the following compact forms utilizing these results:
\bewt{
\eqsp{
% P part
A_{\rm PV}^{(1)} &\, = 2\alpha_B d_J \times \frac{ 1 + {2} y_m \mathtt{R} \cos\Delta\Phi }{2 + y_m^2 \mathtt{R}^2}, \qquad\qquad A_{\rm PV}^{(2)} = 2 F_A \alpha_B \beta^{\frac32} \cdot \sqrt{\frac{ E_c }{ 3\pi \, \Gamma_{J/\psi\to B\bar B} }} \times \frac{1}{\sqrt{2 + y_m^2 \mathtt{R}^2}},\\ 
% CP part
A_{\rm CPV}^{(1)} &\, =  \frac{ {2} (\beta E_c)^{\frac32} y_m \alpha_B \sin\Delta\Phi}{\sqrt{3\pi \, \Gamma_{J/\psi\to B\bar B}}} H_T  \times \mathcal{F}(\mathtt{R}), \quad A_{\rm CPV}^{(2)} = - \frac{ 2 (\beta E_c)^{\frac32} y_m \alpha_B\bar\alpha_B \cos\Delta\Phi}{3\sqrt{3\pi \, \Gamma_{J/\psi\to B\bar B}}} H_T  \times \mathcal{F}(\mathtt{R}),\label{eq:asymsum}
}}
with
\eqal{
\mathcal{F}(\mathtt{R}) \equiv \frac{\mathtt{R}}{\sqrt{2 + y_m^2 \mathtt{R}^2}},
}
where the assumption of real $G_1$ is adopted as commented earlier, {and for clarification, definitions of relevant symbols are given in eq.\,\eqref{eq:jpsidecayff} for $F_A$ and $H_T$, in eq.\,\eqref{eq:diffdw} for $\alpha_B$, below eq.\,\eqref{eq:todd} for $y_m$ and $E_c$, and in eq.\,\eqref{eq:rdelphi} for $\mathtt{R}$ and $\Delta\Phi$, respectively}. {We comment that the parameters in these expressions all depend on the type of $B$ in the final state of $J/\psi$ decay except for $E_c$ and $d_J$, $\mathtt{R}$ is in short of $\mathtt{R}_B$ for instance. Similar to the hadronic form factors of $J/\psi$ decay, we leave out these superscripts/subscripts to keep the results neat.} It is also interesting to note that, for both P and CP violating asymmetries from the decay of $J/\psi$, \textit{i.e.}, $A_{\rm PV}^{(2)}$ and $A_{\rm CPV}^{(1,2)}$, are inversely proportional to the decay width $\Gamma_{J/\psi\to B\bar B}$. This in turn suggests the possibility of testing the P and the CP symmetries from rare decay processes of $J/\psi$ that inherit similar decay patterns as discussed above, especially given that the loss in statistics for these rare processes could possibly be compensated by the very high luminosity of STCFs. It is noted this enhancement might be eliminated/further enhanced by the other decay parameters such as $\alpha$ and $\mathtt{R}$, and a further investment on this possibility is left for a future work.

\begin{table*}[!htb]\caption{Decay parameters for $e^+e^-\to J/\psi\to B\bar B$, with different final states shown in the first row. {\color{blue}Blue} numbers in this table are from\,\cite{ParticleDataGroup:2022pth}. {\color{brown}Brown} numbers are from table 2 of ref.\,\cite{Dobbs:2014ifa} at $\sqrt s = 3.770$\,GeV and momentum transfer $|Q^2|=14.2\rm\,GeV^2$, assuming $|G_E|=|G_M|$. {\color{purple}Purple} numbers are from the DM2 collaboration in\,\cite{DM2:1987riy}. The remaining numbers in black are from the corresponding references in the first row. Whenever given, the first number for each parameter corresponds to its central value, followed by its uncertainties added in quadrature from the original references. ``(derived)'' means results derived from the discussion in section\,\ref{subsec:g12det} will be used for the calculation of asymmetries.}\label{tab:cpparam}
\begin{ruledtabular}
\begin{tabular}{c|ccccccc}
\text{Parameters} & $\Sigma^+\bar\Sigma^-$\,\cite{BESIII:2023ynq} & $\Sigma^-\bar\Sigma^+$ & $\Sigma^0\bar\Sigma^0$\,\cite{BESIII:2017kqw} & $\Lambda\bar\Lambda$\,\cite{BESIII:2022qax} & $p\bar p$\,\cite{BESIII:2019hdp}  & $\Xi^0\bar \Xi^0$\,\cite{BESIII:2023lkg,BESIII:2023drj} & $\Xi^-\bar \Xi^+$\,\cite{BESIII:2021ypr} \\
\hline
$\sqrt{s}\,(\mathrm{GeV})$           & 2.9000            &     ---    &   $m_{J/\psi}$   & $m_{J/\psi}$  &  $3.0800$ & $m_{J/\psi}$ & $m_{J/\psi}$ \\
{$\alpha_{J/\psi}^B$}                                 & $0.35 \pm 0.23$        &    ---         &   $-0.449 \pm 0.022$   & $0.4748 \pm 0.0038$  & --- & $0.514\pm0.016$ & $0.586\pm0.016$ \\
{$\alpha_B$}                                     & {\color{blue}$-0.982\pm0.14$}         &    {\color{blue}$-0.068\pm0.008$}       &     {\color{purple}$0.22\pm0.31$}   & $0.7519 \pm 0.0043$  & {\color{purple}$0.62\pm0.11$} & $-0.3750\pm0.0038$ & $-0.376\pm0.008$ \\
{$\bar\alpha_B$}                               & {\color{blue}$-0.99\pm0.04$}             &    ---       &   ---   & $0.7559\pm0.0078$  & --- & $-0.3790\pm0.0040$  & $-0.371\pm0.007$ \\
$\Delta \Phi\left({\rm\,rad.}\right)$    & $1.3614\pm0.4149$ &    ---       &    ---    & $0.7521\pm0.0066$  & --- & $1.168\pm0.026$  & $1.213\pm0.049$ \\
$\left|{G_E}/{G_M}\right|\equiv\mathtt{R}$                       & $0.8459\pm0.2217$             &    ---        &   {\color{purple}$1.04\pm0.37$}     & $0.8282\pm0.0041$ & {\color{purple}$0.80\pm0.15$} & $0.6672\pm0.0145$  & $0.5986\pm0.0146$ \\
$\left|G_M\right|(\times10^{-2})$                       & (derived)             &    ---        &   {\color{brown}$0.71\pm0.09$}    & (derived) & $3.47\pm0.18$ & (derived)  & (derived) \\
\end{tabular}
\end{ruledtabular}
\end{table*}

%%%%%%%%%%%%%%%%%%%%%
\section{Results}\label{sec:res}
%%%%%%%%%%%%%%%%%%%%%
With the form factors derived in last section, we present our results for testing P and CP symmetries in this section based on the most precise measurements as reported by the BESIII collaboration, as well as the prospects at STCFs for $J/\psi$ decaying into the octet baryon pair final states. 

%%%%%%%
\subsection{Experimental inputs}
%%%%%%%
The experimental inputs we use to estimate the asymmetries are summarized in table\,\ref{tab:cpparam}, where the numbers in black are from BESIII as referenced in the first row, those in blue are instead taken from\,\cite{ParticleDataGroup:2022pth}, in brown from\,\cite{Dobbs:2014ifa} and in purple from\,\cite{DM2:1987riy} due to their limited statistics at BESIII. In particular, for $\Sigma^+$ and $p$,\footnote{Results for $\Sigma^+$ can also be found in\,\cite{BESIII:2020uqk}, which is consistent with those in table\,\ref{tab:cpparam}. We choose the latest results in\,\cite{BESIII:2023ynq}, which will not modify our conclusion of this work. This strategy in data selection is similarly adopted for $p$ in\,\cite{Patterion:2017dzv,Larin:2021sne,BESIII:2021rqk}.} we select numbers in the original references for $\sqrt s$ close to $m_{J/\psi}$. In table\,\ref{tab:cpparam}, whenever given, the first number corresponds to its central value, followed by its uncertainties obtained by adding the reported systematical and statistical ones in quadrature. In particular, the current precision of these measurements are dominantly statistical and will thus be further reduced at STCFs.\footnote{We also comment that for the proton $p$, the value $\mathtt{R}=0.47\pm0.45$ as reported by the BESIII collaboration\,\cite{BESIII:2019hdp} is consistent with that from the DM2 collaboration in\,\cite{DM2:1987riy} but with a larger uncertainty. For this reason, the result in\,\cite{DM2:1987riy} is chosen for presentation in table\,\ref{tab:cpparam}. In addition, the CP parameters of $J/\psi\to\Lambda\bar\Sigma/\bar\Lambda\Sigma^0$ are recently reported in\,\cite{BESIII:2023cvk}.}

%%%%%%%
\subsection{Results for testing P and CP symmetries}\label{subsec:psym}
%%%%%%%
Using measurements summarized in table\,\ref{tab:cpparam}, the form factors discussed in last section and the asymmetries in eq.\,\eqref{eq:asymsum}, $A_{\rm PV}^{(1,2)}$ and the baryon EDMs can be directly obtained for $J/\psi\to B\bar B$ with $B=\Lambda$, $\Sigma^+$, and $\Xi^{0,-}$ where the necessary information is known at this stage. While the current precision is limited by statistics, the STCFs are expected to improve the precision significantly with their luminosities and different center-of-mass energy runs. In particular, a complete measurement of the decay parameters of $\Sigma^{0,-}$ may become possible, the full information of which is now missing due to limited statistics.

\begin{table*}[htb]\caption{P violating asymmetries and baryon EDMs from current measurements summarized in table\,\ref{tab:cpparam}. For the latter, $d_B^{(1)}$ ($d_B^{(2)}$) corresponds to the upper bounds at 95\% CL resulted from $A_{\rm CPV}^{(1)}$ ($A_{\rm CPV}^{(2)}$), assuming statistics dominates the uncertainties at BESIII/STCFs. The last two columns show the statistical uncertainties $\delta$ with 10 billion events from a 12-year running time for BESIII and $t=1$ for one-year data collection at STCF\,\cite{Achasov:2023gey}, assuming the systematical errors are well under control and a detector efficiency of $\epsilon$ indicated in the first column\,\cite{Schonning:2023mge}.}\label{tab:pcpvio}
\begin{ruledtabular}
\begin{tabular}{r|cc|cc|cc|cc}
\multirow{2}{*}{\text{P/CP violation}}   & \multirow{2}{*}{$A_{\rm PV}^{(1)}\,(\times10^{-4})$}    &    \multirow{2}{*}{$A_{\rm PV}^{(2)}\,(\times10^{-4})$}    &   \multicolumn{2}{c|}{${\sqrt{\epsilon\cdot t}}\cdot d_B^{(1)}\,(\times10^{-18}\,e{\rm\,cm})$}   & \multicolumn{2}{c|}{${\sqrt{\epsilon\cdot t}}\cdot d_B^{(2)}\,(\times10^{-18}\,e{\rm\,cm})$}  & \multicolumn{2}{c}{${\sqrt{\epsilon\cdot t}}\cdot{\delta}\,(\times10^{-4})$} \\
\cline{4-9}
                                               &              &               & BESIII   & STCF   & BESIII & STCF & BESIII & STCF      \\
\hline
$\Lambda\,(\epsilon=0.4)$     & {2.88}      & 5.49       & {2.62}      & {0.14}      &  8.64  & 0.47    & 2.30    & 0.13 \\
$\Sigma^+\,(\epsilon=0.2)$    & {$-2.49$} & 7.78       & {1.47}      & {0.08}      & 18.4   & 1.00    & 3.06    & 0.17 \\
$\Xi^0\,(\epsilon=0.2)$           & {$-1.13$} & $-3.45$  & {6.12}      & {0.33}      & 82.6   & 4.41   & 2.92    & 0.16  \\
$\Xi^-\,(\epsilon=0.2)$            & {$-1.08$} & $-2.75$  & {6.79}      & {0.37}      & 95.9   & 5.20   & 3.21    & 0.17 \\
\end{tabular}
\end{ruledtabular}
\end{table*}

The results for $A_{\rm PV}^{(1,2)}$ and the baryon EDMs are summarized in table\,\ref{tab:pcpvio}, where the 95\% CL upper bounds on $d_B^{(1)}$ ($d_B^{(2)}$) are obtained from $A_{\rm CPV}^{(1)}$ ($A_{\rm CPV}^{(2)}$) assuming the statistical uncertainty dominates at BESIII/STCFs with vanishing central values and a Gaussian distribution. Several comments are in order:
\begin{itemize}
\item The P violating asymmetry $A_{\rm PV}^{(1)}$ extracted from the production side of $J/\psi$ is close to but about a factor of two smaller than that extracted from the decay of $J/\psi$, \textit{i.e.}, $A_{\rm PV}^{(2)}$, for $\Lambda$, $\Sigma^+$, and $\Xi^{0,-}$. As a result, $A_{\rm PV}^{(2)}$ is practically more convenient in testing the P symmetry. However, given the detector efficiency, one realizes that both the two P violating asymmetries are not yet measurable at BESIII due to the limited statistics. This situation will change at the next-generation STCFs as indicated in the last column, where all these four channels can be utilized to test the weak theory from P violation.
\item For testing the CP symmetry, we find the baryon EDMs as resulted from $A_{\rm CPV}^{(1)}$ by projecting the momenta of the secondary baryons along the direction of the primary baryons from $J/\psi$ decay, \textit{i.e.}, $d_B^{(1)}$, lead to a factor of a few or even one order of magnitude stronger upper bounds compared with $d_B^{(2)}$ from $A_{\rm CPV}^{(2)}$ as measured by projecting the primary baryon direction onto the decay plane of the secondary baryons. Recall that, as discussed in section\,\ref{sec:setup}, the measurement of $A_{\rm CPV}^{(1)}$ is technically relatively simpler as it does not require a simultaneous reconstruction of the secondary baryon momenta in contrast with that of $A_{\rm CPV}^{(2)}$, $A_{\rm CPV}^{(1)}$ would thus be favored for practical convenience in testing the CP symmetry and measuring the baryon EDMs. On the other hand, the weaker bound on $d_B^{(2)}$, as a consequence of a relative smaller $A_{\rm CPV}^{(2)}$, renders $A_{\rm CPV}^{(2)}$ more sensitive to new CP violating sources beyond the SM that are indispensable for our understanding on the matter-antimatter asymmetry. We also notice that the estimated 95\% CL upper bounds on $d_\Lambda$ at BESIII is already about two orders of magnitude stronger than that in\,\cite{Pondrom:1981gu} from Fermilab.\footnote{An earlier measurement was presented in\,\cite{Baroni:1971uz}, but the constraint is weaker.} This upper bound can be further improved by about one more order of magnitude with just one-year data collection at STCFs, which is also consistent with a recent analysis in\,\cite{Fu:2023ose}. For the other baryons, the upper bounds on their EDMs are of the same order as that of $d_\Lambda$, but slightly above those predicted directly within the quark model as summarized in the ``Reduced Results'' columns in table\,\ref{tab:qedm2dB}.{\color{black} We emphasis that at present there are no direct measurements of $\Xi$ and $\Sigma$ EDMs as can be seen in table\,\ref{tab:pcpvio}. The method discussed here will render their direct measurements possible.}
\item From eq.\,\eqref{eq:asymsum}, the P violating asymmetry $A_{\rm PV}^{(2)}$ is a function of the hadronic decay parameter $\alpha$, $\mathtt{R}$, and the axial-vector form factor $F_A$ where itself depends on the low-energy constants ($D$, $F$), the CKM matrix elements $V_{cd,\, cs}$, the number of colors $N_c$, and the weak mixing angle $\theta_w$. As a consequence, by measuring $A_{\rm PV}^{(2)}$ for these octet baryons, together with their correlations, it would be possible to perform a global analysis based on these asymmetries for a global determination of these parameters. The implication from this global analysis could in turn be used for both the unitary test of the second row of the CKM matrix, and the test of the $\chi$PT/SU(3)$_{L,\,R}$ symmetry. While it remains unclear at this stage how significant the implication would be, we encourage our experimental colleagues in performing such an analysis, especially for the STCFs due to the foreseen unprecedented precision reach.
\item The CP violating asymmetries can also be applied in constraining new CP violating sources. Given the tininess of the baryon EDMs as predicted in the SM, a model-independent approach can be adopted within the SMEFT. In this framework, contact CP violating operators can be interpreted as the sources of baryon EDMs, see for example ref.\,\cite{Engel:2013lsa} for a thorough review. As discussed earlier in section\,\ref{subsec:htres} and summarized in table\,\ref{tab:qedm2dB}, our decomposition of the baryon EDMs in terms of the quark EDMs and CDMs can be understood as the dipole operators in this SMEFT framework. Subject to corrections, for example, from field redefinition and Yukawa modifications which we ignore in the following, these dipole operators are:
\eqsp{
Q_{fW} = &\, \left\{\begin{array}{cl}
(\bar{\mathtt{F}}\sigma^{\mu\nu}f_R)\tau^I{\varphi}W_{\mu\nu}^I, & T_3^{{\mathtt{F}}}=-\frac12\\
(\bar{\mathtt{F}}\sigma^{\mu\nu}f_R)\tau^I\widetilde{\varphi}W_{\mu\nu}^I, & T_3^{{\mathtt{F}}}=\frac12
\end{array}
\right.,\\
Q_{fB} = &\, \left\{\begin{array}{cl}
(\bar{\mathtt{F}}\sigma^{\mu\nu}f_R){\varphi}B_{\mu\nu}, & T_3^{{\mathtt{F}}}=-\frac12\\
(\bar{\mathtt{F}}\sigma^{\mu\nu}f_R)\widetilde{\varphi}B_{\mu\nu}, & T_3^{{\mathtt{F}}}=\frac12
\end{array}
\right.,\\
Q_{uG} = &\, (\bar{Q}\sigma^{\mu\nu}T^Au_R)\widetilde{\varphi}G_{\mu\nu}^A,\\
Q_{dG} = &\, (\bar{Q}\sigma^{\mu\nu}T^Ad_R){\varphi}G_{\mu\nu}^A,
}
with ${\mathtt{F}}$ the SM fermion fields and $f_R$ the corresponding SU(2)$_L$ singlet fermions, $I$ the SU(2)$_L$ index, $u_R$ the up-type and $d_R$ the down-type SU(2)$_L$ singlet quarks. $\varphi$ is the Higgs doublet with $\widetilde\varphi=i\sigma_2\varphi^\ast$ and $\sigma_2$ the second Pauli matrix. We then parameterize the dipole Lagrangian as
\eqsp{
\mathcal{L}_{\rm dipole} = &\,c_{fB} g_1 Q_{fB} + c_{fW} g_2 Q_{fW}\\
&\, + c_{uG} g_3 Q_{uG} + c_{dG} g_3 Q_{dG} + {\rm h.c.}, 
}
where $g_1$, $g_2$ and $g_3$ are, respectively, the U(1)$_Y$, SU(2)$_L$ and SU(3)$_c$ gauge couplings. $c_{fW,fB,uG,dG}$ are Wilson coefficients with mass dimension $-2$, and ${\rm h.c.}$ stands for hermitian conjugate. Then after the spontaneous symmetry breaking and by matching onto the quark EDM and CDM operators as parameterized in eqs.\,\eqref{eq:qedm} and \eqref{eq:qcdm}, one readily finds the quark EDMs $d_q$ and CDMs $f_q$ as given by
\eqsp{
d_q = &\,-\sqrt{2} v\, e\, {\rm Im}\left(c_{qB} - c_{qW}\right),\\
f_q = &\,-\sqrt{2}v\, {\rm Im}\left( c_{qG} \right),
}
with $v\simeq246\rm\,GeV$ the Higgs vacuum expectation value. The strongest upper bounds from $d_\Lambda^{(1)}$ as summarized in table\,\ref{tab:pcpvio} would then imply, as estimated in the quark model summarized in table\,\ref{tab:qedm2dB},
\iffalse
\eqsp{
&\, \left\vert {\rm Im}\left(c_{sB} - c_{sW}\right) \right\vert \lesssim {0.60}\, [{\rm\,TeV}^{-2}], \quad \left\vert {\rm Im}\left(c_{sG}\right) \right\vert \lesssim {1.81}\, [{\rm\,TeV}^{-2}],\qquad \text{(For BESIII)}\\
&\, \left\vert {\rm Im}\left(c_{sB} - c_{sW}\right) \right\vert \lesssim {0.03}\, [{\rm\,TeV}^{-2}], \quad \left\vert {\rm Im}\left(c_{sG}\right) \right\vert \lesssim {0.10}\, [{\rm\,TeV}^{-2}],\qquad \text{(For STCFs)}\\
}
\fi
\eqsp{
&\, \begin{array}{l}
\left\vert {\rm Im}\left(c_{sB} - c_{sW}\right) \right\vert \lesssim {0.60}\, [{\rm\,TeV}^{-2}], \\
\left\vert {\rm Im}\left(c_{sG}\right) \right\vert \lesssim {1.81}\, [{\rm\,TeV}^{-2}],
\end{array}\,\, \text{(For BESIII)}\\
&\, \begin{array}{l}
\left\vert {\rm Im}\left(c_{sB} - c_{sW}\right) \right\vert \lesssim {0.03}\, [{\rm\,TeV}^{-2}], \\
\left\vert {\rm Im}\left(c_{sG}\right) \right\vert \lesssim {0.10}\, [{\rm\,TeV}^{-2}],
\end{array} \,\, \text{(For STCF)}\\
}
corresponding to the sensitivity to new physics around the TeV scale for $\mathcal{O}(1)$ Wilson coefficients.
\end{itemize}

%%%%%%%
\subsection{Weak mixing angle determination}\label{subsec:weakangle}
%%%%%%%
The P violating asymmetries $A_{\rm PV}^{(1,\,2)}$ from the production and decay of $J/\psi$ are sensitive to the weak mixing angle $\theta_w$ as seen in eq.\,\eqref{eq:farge} and \,\eqref{eq:djres}, whose precision measurements have played a key role in both the development and the subsequent precision tests of the SM. Therefore, we think it interesting to investigate the precision to which this parameter can be determined with these P violating asymmetries from $J/\psi$ production and decay, which will be our focus of this subsection.

\begin{table*}[htb]\caption{Precision determination of the weak mixing angle at BESIII using $A_{\rm PV}^{(1)}$ from the production of $J/\psi$, along with the components of its uncertainty.}\label{tab:swunc1}
%\centering{
\begin{ruledtabular}
\begin{tabular}{c|cc|cc|cc|cc}
\text{Baryons}  & $\mathtt{a}_0\,(\times10^{-2})$  & $\frac{\delta m_B}{m_B}\,(\times10^{-6})$ & $\mathtt{a}_1$  & $\frac{\delta m_{J/\psi}}{m_{J/\psi}}\,(\times10^{-6})$    &    $\mathtt{a}_2\,(\times10^{-2})$  & $\frac{\delta\mathtt{R}}{\mathtt{R}}$  & $\mathtt{a}_3$  & $\frac{\delta{\alpha}}{\alpha}\,(\times10^{-2})$ \\
\hline
$\Lambda$     & {9.67}  & 5.38  & {1.09} &  1.94   & {9.67}   &  0.005  & 0.59 &  0.34    \\
$\Sigma^+$    & {8.16}  & 58.9  & {1.27} &  1.94   & {8.16}   &  0.262  & 0.59 &  14.3    \\
$\Xi^0$           & {1.83}  & 152   & {1.17} &  1.94   & {1.83}   &  0.022  & 0.59 &  1.46    \\
$\Xi^-$            & {1.93}  & 53.0  & {1.16} &  1.94   & {1.93}   &  0.024  & 0.59 &  2.85    \\
\hline
\text{(continued)}  & $\mathtt{a}_4$ & $\frac{\delta\Delta\Phi}{\Delta\Phi}\,(\times10^{-2})$ & $\mathtt{a}_5$  & $\frac{\sqrt\epsilon\delta A_{\rm PV}^{(1)}}{A_{\rm PV}^{(1)}}$ & \multicolumn{4}{c}{$(s_w^2 \pm \delta s_w^2)_{\rm BESIII}$} \\
\hline
$\Lambda$     & {0.19}          &  0.88  & 0.59  &  0.52  & \multicolumn{4}{c}{$0.2355\pm{0.1760}$} \\
$\Sigma^+$    & {0.81}          &  30.5  & 0.59  &  1.01  & \multicolumn{4}{c}{$0.2355\pm{0.3884}$} \\
$\Xi^0$           & {0.50}          &  2.23  & 0.59  &  1.88  & \multicolumn{4}{c}{$0.2355\pm{0.8077}$} \\
$\Xi^-$            & {0.51}          &  4.04  & 0.59  &  2.22  & \multicolumn{4}{c}{$0.2355\pm{0.9277}$} \\
\end{tabular}
\end{ruledtabular}
\end{table*}

Due to the simpler form, we start with $A_{\rm PV}^{(1)}$ by taking $(\alpha_{\rm EM},\,m_Z,\,G_F)$ as our inputs and ignoring their uncertainties due to the precision. The uncertainty in $s_w^2$ in this case can then be decomposed as follows
\eqsp{
\frac{\delta s_w^2}{s_w^2} =  &\, \mathtt{a}_0 \frac{\delta m_B}{m_B} \oplus \mathtt{a}_1\, \frac{\delta m_{J/\psi}}{m_{J/\psi}} \oplus \mathtt{a}_2\, \frac{\delta \mathtt{R}}{\mathtt{R}}\\
&\, \oplus \mathtt{a}_3\, \frac{\delta \alpha}{\alpha} \oplus \mathtt{a}_4\, \frac{\delta \Delta\Phi}{\Delta\Phi} \oplus \mathtt{a}_5 \frac{\delta A_{\rm PV}^{(1)}}{A_{\rm PV}^{(1)}},
}
with $\delta x$ the absolute uncertainty in $x$ and the coefficients of each contributing uncertainty summarized in table\,\ref{tab:swunc1}. The absolute uncertainties in $s_w^2$ for each baryon final state can then be calculated by adding all the uncertainties in quadrature with the experimental inputs summarized in table\,\ref{tab:cpparam}. For the asymmetries, as commented above, we take their respective statistical uncertainties as dominant. The results are then shown in the last column, along with the decomposition of $\delta s_w^2$ into its components. Recall that since we use $s_w^2$ at $m_{J/\psi}$ as our benchmark, the central values of $s_w^2$ from different decay channels all align with the SM prediction $s_w^2(m_{J/\psi}) = 0.2355$. Clearly, it is seen from table\,\ref{tab:cpparam} that the uncertainties in $s_w^2$ from both $\delta m_B$ and $\delta m_{J/\psi}$ are generically negligible. In contrast, uncertainties from the other sources, \textit{viz}, the baryon decay parameters and the P violating asymmetries, are comparable and dominate the final uncertainty in $s_w^2$, which are all expected to be reduced at STCFs. The current precision of these decay parameters and the asymmetries hinders the determination $s_w^2$, whose absolute uncertainty is found to be quite large, with the best result $\delta s_w^2 = 0.1760$ from the $\Lambda$ channel.

\begin{figure}
\centering
\includegraphics[width=\columnwidth]{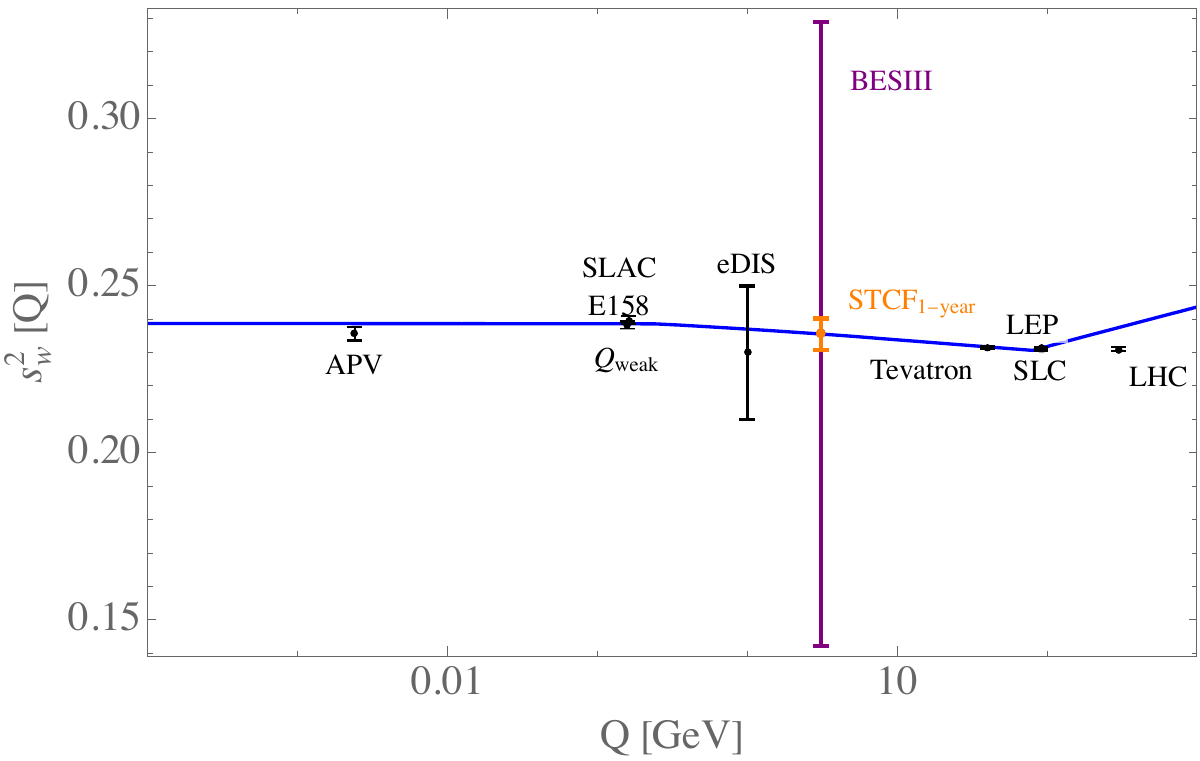}
\caption{Weak mixing angle determination from different experiments. The blue curve represents the SM prediction of the running of the $s_w^2$ in\,\cite{Erler:2004in}. Black data points are from existing experiments as indicated by the legends, and the result from BESIII as obtained in this work is shown in purple with its central value from\,\cite{Erler:2004in} as a benchmark, and similarly for the prospect of STCFs as shown in orange. $s_w^2$ from Tevatron and the LHC are both obtained at the $Z$ pole but shifted horizontally for clarity as in literature.}\label{fig:swdet}
\end{figure}

A similar analysis can be performed using $A_{\rm PV}^{(2)}$, as well as a combined analysis of $A_{\rm PV}^{(1)}$ and $A_{\rm PV}^{(2)}$ fully utilizing the data from both the production and the decay sides of $J/\psi$. Both these two cases are investigated and we refer the readers to appendix\,\ref{app:weak} for details. In each case, new parametric uncertainties enter the total error in $s_w^2$, those for the low-energy constants $D$ and $F$ for instance. With $A_{\rm PV}^{(2)}$ alone and as can be seen in table\,\ref{tab:swunc2}, we find the uncertainties in $s_w^2$ comparable but slightly smaller than the corresponding ones from $A_{\rm PV}^{(1)}$, partially due to the enhancement from relatively larger $A_{\rm PV}^{(2)}$'s. On the other hand, a combined analysis with both $A_{\rm PV}^{(1)}$ and $A_{\rm PV}^{(2)}$, as shown in table\,\ref{tab:swunc3}, leads to $\delta s_w^2={0.09}$ as the most precise determination of $s_w^2$ from the $\Lambda$ baryon decay channel compared with all the other processes considered in this work, corresponding to a relative determination at the $\sim{40\%}$ level. Though not yet competitive with the per mille level precision reach of the next-generation dedicated low-energy experiments such as the MOLLER experiment at the Jefferson Lab\,\cite{MOLLER:2014iki} and the P2 experiment at Mainz\,\cite{Berger:2015aaa}, this measurement at BESIII provides an independent test of the SM at the $J/\psi$ threshold, where our analysis provides a guidance for devoting efforts in reducing the dominant uncertainties.\footnote{It is also noted that a comparable precision with MOLLER/P2 could be achieved at the STCFs using the left-right asymmetry with 80\% beam polarization\,\cite{Bondar:2019zgm,Fu:2023ose}.} Furthermore, at STCFs, the {sensitivity} in $s_w^2$ is expected to be further improved by a factor of about 20 within just one year based on rescaling. For comparison, this determination of $s_w^2$ from $J/\psi\to\Lambda\bar\Lambda$ at BESIII and its prospect at STCFs are also pictorially shown in figure\,\ref{fig:swdet} respectively in purple and orange, with the blue curve the SM prediction from\,\cite{Erler:2004in} and the experimental data points in black taken from the summary report in\,\cite{ParticleDataGroup:2022pth}.

%%%%%%%%%%%%%%%%%%%%%
\section{Conclusions}\label{sec:con}
%%%%%%%%%%%%%%%%%%%%%
In this work, we focus on threshold production of $J/\psi$ at unpolarized electron-positron colliders and its subsequent decay into the lowest-lying baryons for {systematically} testing the P and the CP symmetries. Since the production and decay of $J/\psi$ occur at a very low energy scale compared with $m_Z$, the large logarithms resulted from the large energy gap between the weak scale and the $J/\psi$ decay scale are resummed {by solving the renormalization group equations up to $\mathcal{O}(\alpha_s)$. The effects from the running are found significant, and are also quantitatively summarized in table\,\ref{tab:fanumres} for all the lowest-lying baryons. As a consequence, the axial-vector form factor are either reduced or enhanced by a factor of a few, in addition with sign changes in some cases such as $F_A^{\Xi^0}$ and $F_A^{\Lambda}$. In paticular, $F_A^{\Sigma^0}$ is remarkably reduced by a factor of about $40$ such that its magnitude is two to three orders of magnitude smaller than the other octet baryons, rendering it challenging as a candidate for testing the P symmetry unless enhanced by new mechanism(s) beyond the SM.}

With all the form factors and the decay parameters of $\Lambda$, $\Sigma^+$, and $\Xi^{0,-}$ as reported by the BESIII collaboration using the 10 billion $J/\psi$ events collected at BESIII up to now, we find both P violating asymmetries $A_{\rm PV}^{(1,\,2)}$ of $\mathcal{O}(10^{-4})$. These asymmetries are currently slightly smaller than their statistical uncertainties such that they are not yet observable. This will change with the increased luminosity of STCFs as seen in the last column of table\,\ref{tab:pcpvio}, where the statistical uncertainties are expected to be reduced by at least a factor of 10. On the other hand, as presented in the third and fourth columns of table\,\ref{tab:pcpvio}, the 95\% CL upper bounds on the EDMs of the octet baryons are found to be of $\mathcal{O}(10^{-18})\,e\rm\,cm$ at BESIII. Interestingly, these are already two orders of magnitude stronger than that on $\Lambda$ from Fermilab. Assuming similarly that statistics dominates the uncertainties at STCFs, these upper bounds on the baryon EDMs can be further lifted by about a factor of 10 with just one-year data collection, corresponding to three orders of magnitude improvement compared with that from Fermilab. Especially, a direct probe of the $\Xi$ and $\Sigma$ EDMs can also be achieved with the method discussed in this work.

{Finally,} we also estimate the uncertainty in the weak mixing angle $s_w$ from the P violating asymmetries $A_{\rm PV}^{(1,\,2)}$. In each case, $J/\psi\to\Lambda\bar\Lambda$ leads to the most precise determination of $s_w$ compared with the other baryon final states, thanks to the abundant $J/\psi$ production and the large detector efficiency for $\Lambda$ at BESIII. While the absolute uncertainty in $s_w^2$ is $\delta s_w^2\approx0.1$ using either $A_{\rm PV}^{(1)}$ from the production or $A_{\rm PV}^{(2)}$ from the decay of $J/\psi$, a combined analysis using both data sets reduces this uncertainty down to $\delta s_w^2\approx{0.09}$ as summarized in table\,\ref{tab:swunc3} and pictorially shown in figure\,\ref{fig:swdet}. Though this current uncertainty is still large, it provides an independent measurement of $s_w^2$ at the mass pole of $J/\psi$ and {our analysis also provides a direction} for reducing the dominant uncertainties in the determination of $s_w$ in the future. We also expect the {measurement} of $s_w$ to be improved at the next-generation STCFs {to $\delta s_w^2\approx{0.005}$} as a result of higher luminosities. In this regard, the higher luminosities of next-generation STCFs also render it possible to both improve the current precision in determining the decay parameters of $\Lambda$, $\Sigma^+$, and $\Xi^{0,-}$ and the measurements of the other baryons, \textit{i.e.}, $\Sigma^{0,+}$ that are currently {rather} limited by statistics. As a consequence, a global analysis of $J/\psi$ decaying into all the hyperon final states might become possible, along with a unitary test of the second row of the CKM matrix.

\section*{Acknowledgements}
YD is indebted to the committee and participants of the \textit{Workshop on Hyperon Physics}\,\cite{hyperonws} in Huizhou for communications on BESIII and STCF, especially to Prof. Shuangshi Fang for the discussion on detector efficiencies. YD also thanks Prof. Michael Ramsey-Musolf for helpful discussions, and Dr. Chia-Wei Liu for the discussion on the $t$-channel $W$ exchange process and its renormalization group evolution. This work was partially supported by the Fundamental Research Funds for the Central Universities. XGH was also partially supported by NSFC grant under numbers 12375088 and 12090064. The work of J.P. Ma was supported by National Natural Science Foundation of P.R. China (No.12075299, 11821505 and 11935017) and by the Strategic Priority Research Program of Chinese Academy of Sciences, Grant No. XDB34000000. YD was partially supported by NSFC Special Funds for Theoretical Physics under grant number 12347116, by the China Postdoctoral Science Foundation under grant number 2023M732255, and by the Postdoctoral Fellowship Program of CPSF under number GZC20231613. YD also acknowledges support from the Shanghai Super Postdoc Incentive Plan, and the T.D. Lee Postdoctoral Fellowship at the Tsung-Dao Lee Institute, Shanghai Jiao Tong University.

%%%%%%%%%%%%%%%%%%%%%
\appendix

%%%%%%%%%%%%%%%%%%%%%
\section{Coefficients of the spin decomposition}\label{app:spindecom}
%%%%%%%%%%%%%%%%%%%%%
These coefficients in eq.\,\eqref{eq:spindocom} from the spin decomposition were previously computed in\,\cite{He:2022jjc} for on-shell $J/\psi$ production and its subsequent decay into the $\Lambda$ pair. Several typos are found and we show the correct expression in this appendix.
\begin{widetext}
\begin{small}
\eqal{
	a(\omega) &\, = E_c^2\left[ \abs{G_1}^2 (1+\omega^2)+ \abs{G_2}^2 y_m^2 (1-\omega^2) + 8 d_J {\rm Re}\left( F_A G_1^\ast \right) \beta \omega \right.\nb\\
	&\, \left. \qquad\quad + 4E_c^2 \abs{H_T}^2\beta^2 (1-\omega^2) + \abs{F_A}^2 \beta^2 (1+\omega^2) \right],\\
	c_0(\omega) &\, = \frac13 a(\omega) - \frac{16}{3}E_c^4 \abs{H_T}^2 \beta^2 (1-\omega^2),\\
	c_1(\omega) &\, = -4E_c^3 \beta \omega {\rm Re}\left( H_T G_1^\ast \right) - 8d_J E_c^3 \beta^2 {\rm Re}\left( F_A H_T^\ast \right) , \\
	c_2(\omega) &\, = 4 E_c^3 \beta \biggr\{ y_m {\rm Re}\left( H_T G_2^\ast \right) +  \omega^2 {\rm Re}\left[ H_T (G_1-y_m G_2)^\ast \right] \biggr \} + 8 d_J E_c^3 \beta^2\omega {\rm Re}\left( F_A H_T^\ast \right) , \\
	c_3(\omega) &\, = 4 E_c^3 \abs{\hat{\bm p}\times \hat{\bm k}} \beta \left[ 2d_J {\rm Im}\left( H_T G_1^\ast \right) - \beta\omega {\rm Im}\left( F_A H_T^\ast \right) \right], \\
	c_4(\omega) &\, = 2 E_c^2 \left( \abs{G_1}^2 - \abs{F_A}^2 \beta^2 \right),\\
	c_5(\omega) &\, = 2 E_c^2\left\{ \abs{G_1}^2 - y_m^2 \abs{G_2}^2 + \abs{G_1 - y_m G_2}^2\omega^2 + 4 d_J \beta \omega {\rm Re}\left[ F_A (G_1 - y_m G_2)^\ast \right] \right\},\\
	c_6(\omega) &\, = -2E_c^2 \left[ \omega \abs{G_1}^2 - y_m \omega {\rm Re}\left( G_1G_2^\ast \right) - \omega \beta^2 \abs{F_A}^2 - 2 d_J y_m \beta {\rm Re}\left( F_A G_2^\ast \right) \right],\\
	c_7(\omega) &\, = 2E_c^2 \beta \abs{\hat{\bm p}\times \hat{\bm k}} {\rm Im}\left( F_A G_1^\ast \right), \\
	c_8(\omega) &\, = 2E_c^2 \abs{\hat{\bm p}\times \hat{\bm k}} \biggr\{ 2d_J y_m {\rm Im}\left( G_1 G_2^\ast \right) - \beta\omega {\rm Im}\left[ F_A \left(G_1-y_m G_2\right)^\ast \right] \biggr\},\\
	b_{1n}(\omega) &\, = 2 \abs{\hat{\bm p}\times \hat{\bm k}} E_c^2 \left[ \omega y_m {\rm Im}\left( G_1 G_2^\ast \right) + 2 E_c \omega \beta^2 {\rm Re}\left( F_A H_T^\ast \right) + 2 d_J  y_m \beta {\rm Im}\left( F_A G_2^\ast \right) + 4 d_J E_c \beta {\rm Re}\left( H_T G_1^\ast \right) \right],\\
	b_{2n}(\omega) &\, = b_{1n}(\omega) - 8 \abs{\hat{\bm p}\times \hat{\bm k}} \beta E_c^3 \left[ 2d_J {\rm Re}\left( H_T G_1^\ast \right) + \beta\omega {\rm Re}\left( F_A H_T^\ast \right) \right],\\
	b_{1p}(\omega) &\, = 2E_c^2 \biggr\{ 2 y_m d_J {\rm Re}\left( G_1G_2^\ast \right) + \beta \omega \left[ y_m {\rm Re}\left( F_A G_2^\ast \right) + 2E_c {\rm Im}\left(H_TG_1^\ast\right) \right] - 4d_J E_c \beta^2 {\rm Im}\left( F_A H_T^\ast \right) \biggr\}, \\
	b_{2p}(\omega) &\, = 2E_c^2 \biggr\{ 2 y_m d_J {\rm Re}\left( G_1G_2^\ast \right) + \beta \omega \left[ y_m {\rm Re}\left( F_A G_2^\ast \right) - 2E_c {\rm Im}\left(H_TG_1^\ast\right) \right] + 4 d_J E_c \beta^2 {\rm Im}\left( F_A H_T^\ast \right) \biggr\}, \\
	b_{1k}(\omega) &\, = 2E_c^2 \left\{ 2d_J \omega \left( \abs{G_1}^2 - y_m {\rm Re}\left( G_1 G_2^\ast \right) + \beta^2 \left( \abs{F_A}^2 + 2 E_c {\rm Im}\left( F_A H_T^\ast \right) \right) \right) \right.\nb\\
	 &\, \left. \qquad\qquad + \beta\, {\rm Re}\left[ F_A \left( (1+\omega^2) G_1 - \omega^2 y_m G_2 \right)^\ast \right] - 2E_c \beta\, {\rm Im}\left[ H_T \left( \omega^2 G_1 + (1 - \omega^2) y_m G_2 \right)^\ast \right] \right\}, \\
	b_{2k}(\omega) &\, = 2E_c^2 \left\{ 2d_J \omega \left( \abs{G_1}^2 - y_m {\rm Re}\left( G_1 G_2^\ast \right)  + \beta^2 \left( \abs{F_A}^2 - 2 E_c {\rm Im}\left( F_A H_T^\ast \right) \right) \right) \right.\nb\\
	&\, \left. \qquad\qquad  + \beta\, {\rm Re}\left[ F_A \left( (1+\omega^2) G_1 - \omega^2 y_m G_2 \right)^\ast \right] + 2E_c \beta\, {\rm Im}\left[ H_T \left( \omega^2 G_1 + (1 - \omega^2) y_m G_2 \right)^\ast \right] \right\}.
}
\end{small}
\end{widetext}

%%%%%%%%%%%%%%%%%%%%%
\section{Baryon EDMs in the Quark Model}\label{app:dBres}
%%%%%%%%%%%%%%%%%%%%%
As discussed in the main text, we consider the baryon EDMs $d_B$ as sourced from their quark components. In the effective language, the quarks have both electric and chromo dipole moments, which we discuss case by case below.
%%%%%%%%
\subsubsection{{ Contributions from quark EDMs}}
%%%%%%%%
The quark electric dipole moment (EDM) Lagrangian can be parameterized as
\eqal{
\mathcal{L}_q = -\frac{i}{2}d_q \bar q \sigma^{\mu\nu}\gamma_5 q F_{\mu\nu},\label{eq:qedm}
}
with $d_q$ the quark EDM. In this work, we use the quark model to find their contributions to $d_B$, which predicts
\bewt{
\eqsp{
d_p^{\rm qEDM} &\, = \frac{1}{3}(4d_u - d_d), \quad d_n^{\rm qEDM} = \frac{1}{3}(4d_d - d_u), \\
d_{\Sigma^+}^{\rm qEDM} &\, = \frac{1}{3}(4d_u - d_s), \quad d_{\Sigma^0}^{\rm qEDM} = \frac{1}{3}(2d_u + 2d_d - d_s), \quad d_{\Sigma^-}^{\rm qEDM} = \frac{1}{3}(4d_d - d_s), \\
d_{\Xi^0}^{\rm qEDM} &\, = \frac{1}{3}(4d_s - d_u), \quad d_{\Xi^-}^{\rm qEDM} = \frac{1}{3}(4d_s - d_d), \quad d_{\Lambda^0}^{\rm qEDM} = d_s.
}}

%%%%%%%%
\subsubsection{{ Contributions from quark CDMs}}
%%%%%%%%
Similarly, we write the quark chromo dipole moment (CDM) Lagrangian as
\eqal{
-\mathcal{L}_g  = \mathcal{H}_g = g_s f_q \frac12 \bar q \sigma_{\mu\nu}\gamma_5 T^A q G^{\mu\nu}_A,\label{eq:qcdm}
}
with $g_s$ the SU(3)$_c$ gauge coupling, $f_q$ the quark CDM, $T^A$ the Gell-Mann matrices, and $G$ the gluon field. To find its contribution to $d_B$, we utilize the non-relativistic quark model and treat $\mathcal{H}_g$ perturbatively\,\cite{Khriplovich:1981vs,He:1989mbz}. In this approximation, one neglects the ``induced'' gluon fields as well as its time variation in the non-relativistic limit in analogy with classical electrodynamics, and the perturbation part $\mathcal{H}_g$ reduces to
\eqal{
\mathcal{H}_g = - g_s \sum_k f_q \bdsb{\sigma}_k (-\nabla_{\bdsb{x}_k} G_0^A(\bdsb{x}_k)) T_k^A,
}
with $G_0^a$ the time component of the gluon field potential, $\bdsb{\sigma}$ the Pauli matrices, and the summation over the internal degrees of freedom. In this non-relativistic picture, recall that the unperturbed QCD Hamiltonian is given by
\eqal{
\mathcal{H}_0 = \sum_k \left( \frac{p_k^2}{2m} + g_s T_k^A G_0^A(\bdsb{x}_k) \right),
}
one can readily find
\eqal{
\mathcal{H}_g = i \sum_k f_q \bdsb{\sigma}_k\cdot [\bdsb{p}_k, \mathcal{H}_0],
}
such that the modified baryon wave function becomes
\eqal{
\ket{\tilde{B}} &\, = \left( 1 +  i \sum\limits_k f_q \bdsb{\sigma}_k\cdot \bdsb{p}_k\right) \ket{B},
}
with $\ket{\tilde{B}}$ and $\ket{B}$ the perturbed and unperturbed wavefunction for $B$. It is then straightforward to calculate $d_B$ from $f_q$ as given below:
\bewt{\eqsp{
d_{n}^{\rm qCDM} &\, = \frac{1}{3} \left( -4Q_d f_d + Q_u f_u \right), \quad 
d_{p}^{\rm qCDM} = \frac{1}{3} \left( -4Q_u f_u + Q_d f_d \right), \\
d_{\Sigma^+}^{\rm qCDM} &\, = \frac{1}{3} \left( -4Q_u f_u + Q_s f_s \right), \quad
d_{\Sigma^0}^{\rm qCDM} = \frac{1}{3} \left( -2Q_u f_u -2Q_d f_d + Q_s f_s \right), \quad
d_{\Sigma^-}^{\rm qCDM} = \frac{1}{3} \left( -4Q_d f_d + Q_s f_s \right),\\
d_{\Xi^0}^{\rm qCDM} &\, = \frac{1}{3} \left( -4Q_s f_s + Q_u f_u \right), \quad
d_{\Xi^-}^{\rm qCDM} = \frac{1}{3} \left( -4Q_s f_s + Q_d f_d \right), \quad
d_{\Lambda^0}^{\rm qCDM} = -Q_s f_s,
}}
with $Q_q$ the electric charge of quark $q$.

\clearpage
%%%%%%%%%%%%%%%%%%%%%
\section{$F_A$ derivation in $\chi$PT}\label{app:fader}
%%%%%%%%%%%%%%%%%%%%%

Under ${\rm SU(3)}_L\otimes {\rm SU(3)}_R$, the lowest-lying baryons form an octet and can be parameterized as\footnote{We follow the notations established in \cite{Scherer:2012xha} in this work.}
\eqsp{
B =&\, \sum_{a=1}^8 \frac{B_a \lambda_a}{\sqrt{2}}\\
=&\, \left(\begin{array}{ccc}
\frac{1}{\sqrt{2}} \Sigma^0+\frac{1}{\sqrt{6}} \Lambda & \Sigma^{+} & p \\
\Sigma^{-} & -\frac{1}{\sqrt{2}} \Sigma^0+\frac{1}{\sqrt{6}} \Lambda & n \\
\Xi^{-} & \Xi^0 & -\frac{2}{\sqrt{6}} \Lambda
\end{array}\right)\label{eq:barmx},
}
and the lowest-order Lagrangian respecting this symmetry is given by
\eqsp{
\mathcal{L}_{MB}^{(1)} = &\, {\rm Tr}\left[ \bar B (i \slashed{D} - M_0) B\right]\\
&\, - \frac D 2 {\rm Tr}\left[ \bar B \gamma^\mu \gamma_5 \left\{ u_\mu, B\right\} \right]\\
&\, - \frac F 2 {\rm Tr}\left[ \bar B \gamma^\mu \gamma_5 \left[ u_\mu, B\right] \right],\label{eq:lagmb}
}
with $F$ and $D$ constants determined from semi-leptonic $B$ decays. From tree-level fitting, it was found that $F=0.50$ and $D=0.80$\,\cite{Borasoy:1998pe}, and a recent next-to-leading order analysis up to $\mathcal{O}(p^3)$ in $\chi$PT obtained $F=0.441(47)(2)$ and $D=0.623(61)(17)$\,\cite{Ledwig:2014rfa}, the latter of which will be used in this work. The unitary square root $u$, the chiral connection $\Gamma_\mu$, the chiral vielbein $u_\mu$, and the baryon covariant derivative $D_\mu B$, are defined, respectively, as
\eqsp{
&\, u = U^{\frac12} = \exp\left(i\frac{\phi}{2F_0}\right),\\
&\, \Gamma_\mu = \frac12\left[ u^\dagger\left( \partial_\mu - ir_\mu \right)u + u\left( \partial_\mu - i l_\mu \right)u^\dagger \right],\\
&\, u_\mu \equiv i \left[ u^\dagger\left( \partial_\mu - ir_\mu \right)u - u\left( \partial_\mu - i l_\mu \right)u^\dagger \right],\\
&\, D_\mu B = \partial_\mu B + [\Gamma_\mu, B],
}
with $F_0\approx F_\pi = 92.4\rm\,MeV$ and $F_\pi$ the pion decay constant, and $U$ the transformation matrix for the meson fields $\phi$ parameterized as
\eqsp{
\phi =&\, \sum_{a=1}^8 \phi_a \lambda_a\\
= &\, \left(\begin{array}{ccc}
\pi^0 + \frac{1}{\sqrt3}\eta & \sqrt2\pi^+ & \sqrt2 K^+ \\
\sqrt2\pi^- & -\pi^0 + \frac{1}{\sqrt3}\eta & \sqrt2 K^0 \\
\sqrt2 K^- & \sqrt2\bar K^0 & -\frac{2}{\sqrt3}\eta
\end{array}\right).
}
$r_\mu$ and $l_\mu$ account for the external sources. Plugging the quark currents as represented by terms inside the square bracket of eq.\,\eqref{eq:faextract} into eq.\,\eqref{eq:lagmb} and then matching with eq.\,\eqref{eq:jpsidecayff}, one obtains results in eq.\,\eqref{eq:farge}.

\begin{table*}[htb]\caption{Precision determination of the weak mixing angle at BESIII using $A_{\rm PV}^{(2)}$ from the decay of $J/\psi\to B\bar B$ with $B=\Lambda,\Sigma^{+},\Xi^{0,-}$, along with the components of its uncertainty.}\label{tab:swunc2}
\begin{ruledtabular}
\begin{tabular}{c|cc|cc|cc|cc|cc|cc}
\text{Baryons}   & $\mathtt{b}_0$  & $\frac{\delta m_B}{m_B}\,(\times10^{-6})$ & $\mathtt{b}_1$ & $\frac{\delta m_{J/\psi}}{m_{J/\psi}}\,(\times10^{-6})$ &    $\mathtt{b}_2$ &  $\frac{\delta\mathtt{R}}{\mathtt{R}}$  &   $\mathtt{b}_3$ & $\frac{\delta{\alpha}}{\alpha}\,(\times10^{-2})$ & $\mathtt{b}_4$ & $\frac{\delta V_{cs}}{V_{cs}}$ & $\mathtt{b}_5$  & $\frac{\delta {\Gamma_{J/\Psi\to B\bar B}}}{{\Gamma_{J/\Psi\to B\bar B}}}$\\
\hline
$\Lambda$     & 1.21  &  5.38  & 1.55  &  1.94  & 0.10   &  0.005  & 0.68  & 0.34   & 0.14     &  0.011    & 0.34  &  0.048 \\
$\Sigma^+$    & 1.38  &  58.9  & 1.68  &  1.94  & 0.10   &  0.262  & 0.59  & 14.3   & 0.04     &  0.011    & 0.30  &  0.037 \\
$\Xi^0$           & 2.69  &  152   & 3.02  &  1.94  & 0.09   &  0.022  & 0.67  & 1.46   & 0.14     &  0.011    & 0.33  &  0.034 \\
$\Xi^-$            & 2.90  &  53.0  & 3.25  &  1.94  & 0.08   &  0.024  & 0.70  & 2.85   & 0.23     &  0.011    & 0.35  &  0.082 \\
\hline
\text{(continued)}   & $\mathtt{b}_6$ & $\frac{\sqrt\epsilon\delta A_{\rm PV}^{(2)}}{A_{\rm PV}^{(2)}}$ &    $\mathtt{b}_7\,(\times10^{-2})$ & $\frac{\delta D}{D}$   &   $\mathtt{b}_8$ & $\frac{\delta F}{F}$ & $\mathtt{b}_9$ & $\frac{\delta g_V}{g_V}\,(\times10^{-3})$ & \multicolumn{4}{c}{$(s_w^2\pm\delta s_w^2)_{\rm BESIII}$} \\
\hline
$\Lambda$     & 0.51  &  0.42  &  68.1  &  0.10  & 0       &  0.11  &  0.68    &  9.59   & \multicolumn{4}{c}{$0.2355\pm0.1058$}\\
$\Sigma^+$    & 0.57  &  0.39  &  0.07  &  0.10  & 0.59  &  0.11  &  0.59    &  9.59   & \multicolumn{4}{c}{$0.2355\pm0.1250$} \\
$\Xi^0$           & 0.49  &  0.91  &  63.6  &  0.10  & 0.03  &  0.11  &  0.67    &  9.59   & \multicolumn{4}{c}{$0.2355\pm0.2949$} \\
$\Xi^-$            & 0.49  &  1.26  &  6.52  &  0.10  & 0.64  &  0.11  &  0.70    &  9.59   & \multicolumn{4}{c}{$0.2355\pm0.4231$} \\
\end{tabular}
\end{ruledtabular}
\end{table*}

\begin{table*}[htb]\caption{Precision determination of the weak mixing angle at BESIII from a combined analysis $A_{\rm PV}^{(1,\,2)}$, along with the components of its uncertainty.}\label{tab:swunc3}
\begin{ruledtabular}
\begin{tabular}{c|cc|cc|cc|cc|cc|cc}
\text{Baryons}   & $\mathtt{c}_0$  & $\frac{\delta m_B}{m_B}\,(\times10^{-6})$ & $\mathtt{c}_1$ & $\frac{\delta m_{J/\psi}}{m_{J/\psi}}\,(\times10^{-6})$ &    $\mathtt{c}_2$ &  $\frac{\delta {\Gamma_{J/\Psi\to B\bar B}}}{{\Gamma_{J/\Psi\to B\bar B}}}$  &   $\mathtt{c}_3$ & $\frac{\delta\mathtt{R}}{\mathtt{R}}$ & $\mathtt{c}_4$ & $\frac{\delta{\alpha}}{\alpha}$ & $\mathtt{c}_5$  & $\frac{\delta V_{cs}}{V_{cs}}$ \\
\hline
$\Lambda$     & {0.72}  &  5.38  & {1.38}  &  1.94  & {0.21}   &  0.048  & {0.03}  & 0.005   & {0.65}     &  0.003    & {0.09}  & 0.011  \\
$\Sigma^+$    & {1.99}  &  58.9  & {1.87}  &  1.94  & {0.44}   &  0.037  & {0.11}  & 0.262   & 0.59                 &  0.14    & {0.06}  & 0.011  \\
$\Xi^0$           & {1.96}  &  152   & {2.52}  &  1.94  & {0.25}   &  0.034  & {0.06}  & 0.022   & 0.65                 &  0.01    & {0.10}  & 0.011  \\
$\Xi^-$            & {1.99}  &  53.0  & {2.60}  &  1.94  & {0.24}   &  0.082  & {0.05}  & 0.024   & {0.67}     &  0.03    & {0.16}  & 0.011  \\
\hline
\text{(continued)}   & $\mathtt{c}_6$ & $\frac{\delta D}{D}$ &    $\mathtt{c}_7$ &  $\frac{\delta F}{F}$ &   $\mathtt{c}_8$ & $\frac{\delta \Delta\Phi}{\Delta\Phi}$ & $\mathtt{c}_9$ & $\frac{\delta g_V}{g_V}$ & $\mathtt{c}_{10}$ & $\frac{\sqrt\epsilon\delta A_{\rm PV}^{(1)}}{A_{\rm PV}^{(1)}}$ & $\mathtt{c}_{11}$ & $\frac{\sqrt\epsilon\delta A_{\rm PV}^{(2)}}{A_{\rm PV}^{(2)}}$ \\
\hline
$\Lambda$     & {0.43}        &  0.10  &  0                  &  0.11  & {0.07}  &  0.01  &  {0.43}    &  0.01   & {0.22}  &  0.56 & {0.43}  &  0.45  \\
$\Sigma^+$    & 0.001                 &  0.10  &  {0.87}  &  0.11  & {0.38}  &  0.31  &  {0.87}    &  0.01   & {0.28}  &  1.09 & {0.87}  &  0.42  \\
$\Xi^0$           & {0.46}        &  0.10  &  {0.02}  &  0.11  & {0.13}  &  0.02  &  {0.49}    &  0.01   & {0.16}  &  2.02 & {0.49}  &  0.98  \\
$\Xi^-$            & 0.04                   &  0.10  &  {0.44}  &  0.11  & {0.16}  &  0.04  &  {0.48}    &  0.01   & {0.19}  &  2.39 & {0.48}  &  1.36  \\
\hline
{\multirow{2}{*}{$(s_w^2\pm\delta s_w^2)_{\rm BESIII}$}} & \multicolumn{12}{c}{\multirow{2}{*}{$\left(0.2355\pm{0.0933}\right)_{\Lambda} \quad \left(0.2355\pm{0.2574}\right)_{\Sigma^+} \quad \left(0.2355\pm{0.1340}\right)_{\Xi^0} \quad \left(0.2355\pm{0.3054}\right)_{\Xi^-}$}}\\
& \multicolumn{12}{c}{}\\
\end{tabular}
\end{ruledtabular}
\end{table*}

%%%%%%%%%%%%%%%%%%%%%
\section{Weak mixing angle from $A_{\rm PV}^{(2)}$ and the combination of $A_{\rm PV}^{(1,\,2)}$}\label{app:weak}
%%%%%%%%%%%%%%%%%%%%%
Similar to the discussion on weak mixing angle determination with $A_{\rm PV}^{(1)}$, we present our discussion on this point utilizing $A_{\rm PV}^{(2)}$ and a combined analysis of $A_{\rm PV}^{(1,\,2)}$ in the following. For that with $A_{\rm PV}^{(2)}$, we decompose the uncertainty in $s_w$ as follows:
\eqsp{
\frac{\delta s_w^2}{s_w^2} = &\, \mathtt{b}_0 \frac{\delta m_B}{m_B} \oplus \mathtt{b}_1\, \frac{\delta m_{J/\psi}}{m_{J/\psi}} \oplus \mathtt{b}_2 \, \frac{\delta \mathtt{R}}{\mathtt{R}} \oplus \mathtt{b}_3\, \frac{\delta \alpha}{\alpha}\\
&\, \oplus \mathtt{b_4} \, \frac{\delta V_{cs}}{V_{cs}} \oplus \mathtt{b}_5\, \frac{\delta {\Gamma_{J/\Psi\to B\bar B}}}{{\Gamma_{J/\Psi\to B\bar B}}} \oplus \mathtt{b}_6 \frac{\delta A_{\rm PV}^{(2)}}{A_{\rm PV}^{(2)}}\\
&\, \oplus \mathtt{b}_7 \frac{\delta D}{D} \oplus \mathtt{b}_8 \frac{\delta F}{F} \oplus \mathtt{b}_9 \frac{\delta g_V}{g_V},
}
and summarize the decomposition and the final uncertainty in $s_w$ in table\,\ref{tab:swunc2}. Similar to the analysis for $A_{\rm PV}^{(1)}$, it is clear that the total uncertainty in $s_w^2$ is about 0.1 and quite large, dominated by the precision in the decay parameters, the asymmetry $A_{\rm PV}^{(2)}$, and the low-energy constants. While largely rooted in limited statistics at BESIII for the first two sources, the uncertainty in the last one, \textit{i.e.}, the low-energy constant, can be improved either through a lattice calculation, or a global fit to data with more statistics in the future.

The large uncertainty in $s_w$ from either $A_{\rm PV}^{(1)}$ or $A_{\rm PV}^{(2)}$ can be reduced by combining the two asymmetries from the production and the decay of $J/\psi$. In this case, the uncertainty in $s_w^2$ can be decomposed as:
\eqsp{
\frac{\delta s_w^2}{s_w^2} = &\, \mathtt{c}_0 \frac{\delta m_B}{m_B} \oplus \mathtt{c}_1\, \frac{\delta m_{J/\psi}}{m_{J/\psi}} \oplus \mathtt{c}_2 \frac{\delta {\Gamma_{J/\Psi\to B\bar B}}}{{\Gamma_{J/\Psi\to B\bar B}}}\\
&\, \oplus \mathtt{c}_3\, \frac{\delta \mathtt{R}}{\mathtt{R}} \oplus \mathtt{b_4} \, \frac{\delta \alpha}{\alpha} \oplus \mathtt{c}_5 \frac{\delta V_{cs}}{V_{cs}} \oplus \mathtt{c}_6 \frac{\delta D}{D}\\
&\, \oplus \mathtt{c}_7 \frac{\delta F}{F} \oplus \mathtt{c}_8 \frac{\delta \Delta\Phi}{\Delta\Phi} \oplus \mathtt{c}_9 \frac{\delta g_V}{g_V} \oplus \mathtt{c}_{10} \frac{\delta A_{\rm PV}^{(1)}}{A_{\rm PV}^{(1)}}\\
&\, \oplus \mathtt{c}_{11} \frac{\delta A_{\rm PV}^{(2)}}{A_{\rm PV}^{(2)}},
}
with the coefficients and the final uncertainty from different decay channels summarized in table\,\ref{tab:swunc3}. Note that to enhance the asymmetry, we sum over these two asymmetries for $\Lambda$ and $\Xi^{0,-}$ while subtract them for $\Sigma^+$ for this combined analysis. Once again, we find $J/\psi\to\Lambda\bar\Lambda$ leads to the most precise determination of $s_w^2$ with an absolute uncertainty $\delta s_w^2=0.0788$ compared with the other baryon channels.

\clearpage
\bibliography{ref}
\end{document}